\documentclass[12pt]{article} 

\begin{document} 
 
\def\CA{{\cal A}} 
\def\CB{{\cal B}} 
\def\CC{{\cal C}} 
\def\CD{{\cal D}} 
\def\CE{{\cal E}} 
\def\CF{{\cal F}} 
\def\CG{{\cal G}} 
\def\CH{{\cal H}} 
\def\CI{{\cal I}} 
\def\CJ{{\cal J}} 
\def\CK{{\cal K}} 
\def\CL{{\cal L}} 
\def\CM{{\cal M}} 
\def\CN{{\cal N}} 
\def\CO{{\cal O}} 
\def\CP{{\cal P}} 
\def\CQ{{\cal Q}} 
\def\CR{{\cal R}} 
\def\CS{{\cal S}} 
\def\CT{{\cal T}} 
\def\CU{{\cal U}} 
\def\CV{{\cal V}} 
\def\CW{{\cal W}} 
\def\CX{{\cal X}} 
\def\CY{{\cal Y}} 
\def\CZ{{\cal Z}} 
 
\newcommand{\todo}[1]{{\em \small {#1}}\marginpar{$\Longleftarrow$}} 
\newcommand{\labell}[1]{\label{#1}\qquad_{#1}} 
\newcommand{\bbibitem}[1]{\bibitem{#1}\marginpar{#1}} 
\newcommand{\llabel}[1]{\label{#1}\marginpar{#1}} 
 
\newcommand{\sphere}[0]{{\rm S}^3} 
\newcommand{\su}[0]{{\rm SU(2)}} 
\newcommand{\so}[0]{{\rm SO(4)}} 
\newcommand{\bK}[0]{{\bf K}} 
\newcommand{\bL}[0]{{\bf L}} 
\newcommand{\bR}[0]{{\bf R}} 
\newcommand{\tK}[0]{\tilde{K}} 
\newcommand{\tL}[0]{\bar{L}} 
\newcommand{\tR}[0]{\tilde{R}}

\newcommand{\btzm}[0]{BTZ$_{\rm M}$} 
\newcommand{\ads}[1]{{\rm AdS}_{#1}} 
\newcommand{\SL}[0]{{\rm SL}(2,R)} 
\newcommand{\cosm}[0]{R} 
\newcommand{\hdim}[0]{\bar{h}} 
\newcommand{\bw}[0]{\bar{w}} 
\newcommand{\bz}[0]{\bar{z}} 
\newcommand{\be}{\begin{equation}} 
\newcommand{\ee}{\end{equation}} 
\newcommand{\bea}{\begin{eqnarray}} 
\newcommand{\eea}{\end{eqnarray}} 
\newcommand{\pat}{\partial} 
\newcommand{\lp}{\lambda_+} 
\newcommand{\bx}{ {\bf x}} 
\newcommand{\bk}{{\bf k}} 
\newcommand{\bb}{{\bf b}} 
\newcommand{\BB}{{\bf B}} 
\newcommand{\tp}{\tilde{\phi}} 
\hyphenation{Min-kow-ski}

\def\apr{\alpha'} 
\def\str{{str}} 
\def\lstr{\ell_\str} 
\def\gstr{g_\str} 
\def\Mstr{M_\str} 
\def\lpl{\ell_{pl}} 
\def\Mpl{M_{pl}} 
\def\varep{\varepsilon} 
\def\del{\nabla} 
\def\grad{\nabla} 
\def\tr{\hbox{tr}} 
\def\perp{\bot} 
\def\half{\frac{1}{2}} 
\def\p{\partial} 
\def\perp{\bot} 
\def\eps{\epsilon} 
 
\renewcommand{\thepage}{\arabic{page}} 
\setcounter{page}{1}

\rightline{DTP/00/101, UPR-908-T, ITFA-2000-31} 
\rightline{HIP-2000-63/TH, hep-th/0011217} 
\vskip 1cm 
\centerline{\Large \bf Supersymmetric Conical Defects: } 
\centerline{\large \bf Towards a string theoretic description of black hole 
formation} 
\vskip 1cm 
 
\renewcommand{\thefootnote}{\fnsymbol{footnote}} 
\centerline{{\bf Vijay 
Balasubramanian${}^{1}$\footnote{vijay@endive.hep.upenn.edu}, 
Jan de Boer${}^{2}$\footnote{jdeboer@wins.uva.nl}, 
Esko Keski-Vakkuri${}^{3}$\footnote{keskivak@rock.helsinki.fi} 
and 
Simon F. Ross${}^{4}$\footnote{S.F.Ross@durham.ac.uk}}} 
\vskip .5cm 
\centerline{${}^1$\it David Rittenhouse Laboratories, University of 
Pennsylvania,} 
\centerline{\it Philadelphia, PA 19104, USA} 
\vskip .5cm 
\centerline{${}^2$\it Instituut voor Theoretische Fysica,} 
\centerline{\it Valckenierstraat 65, 1018XE Amsterdam, The Netherlands} 
\vskip .5cm 
\centerline{${}^3
$\it Fysiikan tutkimuslaitos, Helsingin Yliopisto,} 
\centerline{\it PL 9, FIN-00014 Helsingin Yliopisto, Finland} 
\vskip .5cm 
\centerline{${}^4$ \it Centre for Particle Theory, Department of 
Mathematical Sciences} 
\centerline{\it University of Durham, South Road, Durham DH1 3LE, U.K.} 
 
\setcounter{footnote}{0} 
\renewcommand{\thefootnote}{\arabic{footnote}} 
 
\begin{abstract} 
Conical defects, or point particles, in $\ads{3}$ are one of the
simplest non-trivial gravitating systems, and are particularly
interesting because black holes can form from their collision.  We
embed the BPS conical defects of three dimensions into the ${\cal N}
=4b$ supergravity in six dimensions, which arises from IIB string
theory compactified on K3.  The required Kaluza-Klein reduction of the
six dimensional theory on a sphere is analyzed in detail, including
the relation to the Chern-Simons supergravities in three dimensions.
We show that the six dimensional spaces obtained by embedding the 3d
conical defects  arise in the near-horizon limit of rotating black
strings.  Various properties of these solutions are analyzed and we
propose a representation of our defects in the CFT dual to
asymptotically $\ads{3} \times S^3$ spaces.  Our work is intended as a
first step towards analyzing colliding defects that form black holes.
\end{abstract}

 
\section{Introduction} 
\label{intro} 
 
Twenty-five years after Hawking showed that black holes emit thermal
radiation~\cite{hawking}, the apparent loss of quantum mechanical
unitarity in the presence of a black hole remains an outstanding
problem for theoretical physics.  We expect that this ``information
puzzle'', which represents a fundamental tension between general
relativity and quantum mechanics, should either be erased or explained
in a quantum theory of gravity.  In recent years string theory has
explained microscopically the huge degeneracy required to account for
the entropy of certain extremal black holes.  However, there has been
no insight into why this degeneracy of states is related to something
geometric like the area of a horizon.  More fundamentally, the
information puzzle remains exactly that -- a puzzle.
 
This paper is the first in a series investigating the black hole
information puzzle in the context of string theory.  In General
Relativity, the simplest context for black hole formation is gravity
in three dimensions where there are no local dynamics.  In the
presence of a negative cosmological constant, 3d gravity possesses
black hole solutions~\cite{btz}.  There is also a family of conical
defects, the so-called point particles~\cite{desjack}.  These
solutions interpolate between the vacuum solution ($\ads{3}$ with mass
$M = -1$ in conventional units) and the black hole spectrum which
starts at $M=0$. Exact solutions of 3d gravity are known in which the
collision of conical defects forms a black hole~\cite{coll}.  We would
like to use these simple classical processes to study the formation of
higher dimensional black holes in string theory. To this end, we must
first embed the conical defects supersymmetrically in a higher
dimensional gravity arising from string theory.  Preserving
supersymmetry is important because the controlled quantization of
black holes and solitons in string theory usually requires
supersymmetry.  The presence of the negative cosmological constant in
three dimensions suggests that there should be a dual description of
such spaces in terms of a two-dimensional conformal field
theory~\cite{juanads}.  Our goal is to find such a dual picture and
describe in it the process of black hole formation from collision of
conical defects.  In~\cite{vijsim} it was shown that the 3d conical
defects and their collisions can be detected in correlation functions
of the dual CFT.  Here we are interested in the direct description of
the defects as objects in the dual.\footnote{It would also be
interesting to make contact with the investigations of spherical
shells in~\cite{DKK}.}

Type IIB supergravity compactified on K3 yields the chiral ${\cal N} = 
4b$ supergravity in six dimensions, coupled to 21 tensor
multiplets.  This theory has classical solutions 
with the geometry of $\ads{3} \times S^3$.  In Sec.~\ref{sec:kk} we 
will construct supersymmetric solutions where the sphere is fibered 
over $\ads{3}$ so that a minimum length circuit around the $\ads{3}$ 
base leads to a rotation of the sphere around an axis.  Since 
$\ads{3}$ is simply connected, the fibre must break down at a point. 
Upon dimensional reduction to the base this produces supersymmetric 
conical defects in three dimensions.  In fact, the identical objects 
have been obtained previously as solutions to extended $2+1$ 
supergravity in the Chern-Simons formulation~\cite{Izq:2+1, 
david:point}.  The $U(1)$ Wilson lines used in these constructions to 
obtain a BPS solution arise in our case from the Kaluza-Klein gauge 
field associated with the fibration.    Our Kaluza-Klein ansatz for
reducing the action and equations of motion of 6d gravity to
 the 3d base does not
yield precisely a Chern-Simons theory. Nevertheless, the dimensionally
reduced system admits solutions with vanishing field strength,
for which the analysis of supersymmetry remains 
unchanged -- the holonomy of Killing spinors under the spin connection 
is cancelled by the holonomy under the gauge connection. Various 
details of sphere compactifications of 6d, ${\cal N} = 4b$ 
supergravity are reviewed in the main text and the
appendices.\footnote{Sphere compactifications have been extensively
studied in the literature.  See~\cite{duffetal} for a review, and the
recent work~\cite{spherecomp} for references.} 
 
It is well-known that a horospheric patch of the $\ads{3} \times S^3$ 
geometry can be obtained as a near-horizon limit of the black string 
soliton of 6d supergravity~\cite{nearhor}.  Compactifying 
the extremal string solution on a circle yields the black holes of 
five dimensional string theory whose states were counted in the 
classic paper~\cite{stromvaf}.\footnote{The $\ads{3}$ and BTZ geometries can 
also be related to the near-horizon limit of extremal four dimensional 
black holes~\cite{bl4d}, by constructing the black holes as the near 
horizon limit of intersecting 5-branes in M-theory.}  The near-horizon 
limit of these solutions yields the BTZ black holes times 
$S^3$~\cite{nearhor}.  In Sec.~\ref{sec:spin} we show that the fibered 
$S^3$ solutions described above arise as the near-horizon geometries 
of an extremal limit of spinning 6-dimensional strings 
compactified on a circle.  Interestingly, when the angular momentum is 
suitably chosen, {\it global} $\ads{3} \times S^3$ is recovered as a 
solution.  We discuss various properties of the solution, including the 
nature of the conical singularity and potential Gregory-Laflamme 
instabilities in the approach to extremality.
 
The near-horizon limit of the six dimensional black string is also a
decoupling limit for the worldvolume CFT description of the soliton.
Following the reasoning of~\cite{juanads} we conclude that the BPS
conical defects described above should enjoy a non-perturbative dual
description in the worldvolume CFT of the black string -- i.e., a
deformation of the orbifold sigma model
$(K3)^N/S^N$~\cite{finnemil}.  When reduced to the AdS base,
the fibered geometries appearing in our solutions carry a U(1) charge
measured by the Wilson line holonomy.  Within the AdS/CFT duality,
this spacetime U(1) charge translates into an R-charge of the dual
system.  In Sec.~\ref{sec:dual}, we propose that the conical defects
are described in the dual as an ensemble of the chiral primaries
carrying the same R-charge. In subsequent papers we will test
this proposal and then use it to analyze the spacetime scattering of
conical defects.

 
\section{Conical defects from Kaluza-Klein reduction} 
\label{sec:kk} 
 
In this section, we obtain the supersymmetric conical 
defects in 3d via Kaluza-Klein reduction of the six-dimensional ${\cal N} = 4b$ 
supergravity.    Defects in three dimensions that involve just the
metric and gauge fields with a Chern-Simons action have been obtained
previously~\cite{Izq:2+1}.   We will  construct a Kaluza-Klein ansatz
for six dimensional gravity which reproduces these defects upon
dimensional reduction.

We begin by reviewing the structure of the 3d conical defects.
The action with a negative cosmological constant is 
\begin{equation} 
S = {1 \over 16\pi G_3} \int_{\cal M} d^3x\, \sqrt{-g} \left( R + {2 
\over \ell^2} \right)   - {1\over 8\pi G_3} \int_{\partial {\cal M}} 
\sqrt{-h} \left( \theta + {1 \over \ell} \right), 
\end{equation} 
where $\theta$ is the trace of the extrinsic curvature of the 
boundary. The boundary term $\sqrt{-h} \, \theta$ renders the 
equations of motion well-defined, leading to the solutions 
\begin{equation} 
ds^2 = -\left( {r^2 \over \ell^2} - M_3 \right) \, dt^2 + 
\left( {r^2 \over \ell^2} - M_3 \right)^{-1} \, dr^2 + r^2 \, d\phi^2 
\ , \label{ppmetric} 
\end{equation} 
where $\phi \sim \phi + 2\pi$. $M_3 = -1$ is the vacuum, global anti-de 
Sitter space ($\ads{3}$).  The boundary term $\sqrt{-h}/\ell$ 
renders the action finite for any solution that approaches the vacuum 
sufficiently rapidly at infinity~\cite{stressvp}.  The mass of these solutions can 
then be computed following~\cite{stressvp,adsstress} to be $M = M_3/8G_3$. 
The $M_3 \geq 0$ solutions are the non-rotating BTZ black 
holes~\cite{btz} while the spacetimes in the range $-1 < M_3 \leq 0$ 
are conical defects~\cite{desjack}.  To display the defect, let 
$\gamma^2 \equiv -M_3$ and rescale the coordinates: $\hat{t} 
\equiv  t \, \gamma$, $\hat{r} \equiv r/\gamma$, and 
$\hat{\phi} = \phi \, \gamma$.   Then 
\begin{equation} 
ds^2 = - \left( 1 + {\hat{r}^2 \over \ell^2} \right) \, d\hat{t}^2 + 
\left( 1 + {\hat{r}^2 \over \ell^2} \right)^{-1} \, d\hat{r}^2 + 
\hat{r}^2 \, d\hat{\phi}^2,   
\end{equation} 
where $\hat{\phi} \sim \hat{\phi} + 
2\pi\gamma$, manifestly exhibiting a deficit angle of 
\begin{equation} 
\delta\hat{\phi} = 2\pi(1 - \gamma). 
\end{equation} 
In these coordinates the mass measured with respect to translations in
$\hat{t}$ is $M = - \sqrt{-M_3}/8  G_3$.    
 
We are looking for an embedding of these solutions in the ${\cal N} 
=4b$ chiral supergravity in six dimensions~\cite{romans},
coupled to tensor multiplets. The theory 
has self-dual tensor fields, so it has solutions where three 
directions are spontaneously compactified on $S^3$; the vacuum for 
this sector is $\ads{3}$, and the spectrum of fluctuations around this 
vacuum solution has been 
computed~\cite{deger:6d,fluc2,janads}. We seek
a supersymmetric solution where $\ads{3}$ is replaced by a
conical defect.  
 
In extended three dimensional supergravity, the conical defects can be
made supersymmetric~\cite{Izq:2+1}.  These BPS defects achieve
supersymmetry by cancelling the holonomy of spinors under the spin
connection by the holonomy under a Wilson line of a flat gauge field
appended to the solution.  Thus, we will consider a Kaluza-Klein
ansatz which involves non-trivial Kaluza-Klein gauge fields (leading to
a fibered $S^3$ in the 6d geometry) and the three dimensional metric,
since these were the only fields present in the extended
three-dimensional supergravities.
 
Famously, three-dimensional gravity can be written as a sum two $\SL$
Chern-Simons theories.  The sphere reduction of six-dimensional,
${\cal N} = 4b$ gravity has symmetries appropriate to the ${\rm
SU(1,1|2)} \times {\rm SU(1,1|2)}$ Chern-Simons supergravity
(see~\cite{AchuTown,hamred2,janads,NT,david:group,david:point} and
references therein). We will show that the three-dimensional equations
of motion obtained from our Kaluza-Klein ansatz contain the (bosonic)
solutions of this theory. However, the six-dimensional action does not
reduce to Chern-Simons in three dimensions. In fact, the equations of
motion obtained from our ansatz are not obtainable from a
three-dimensional action; we would have to include some non-trivial
scalars in our general ansatz to obtain a consistent truncation to a
three-dimensional action. That is, while our ansatz shows that we can
construct solutions of the six-dimensional theory using all the
bosonic solutions of the ${\rm SU(1,1|2)} \times {\rm SU(1,1|2)}$
supergravity, asking that our ansatz solve the six-dimensional
equations does not in general give the equations of motion of a
three-dimensional theory.
 
The minimal ${\cal N} = 4b$ theory contains a graviton $e^A_M$, four 
left-handed gravitini $\psi_{Mr}$, and five antisymmetric tensor fields 
$B^{\mathsf{i}}_{MN}$. The latter transform under the vector representation of 
${\rm Spin}(5)$. We adopt a notation where curved spacetime indices 
are: $M,N = 0 \ldots 5$ for the full six-dimensional geometry; 
$\mu,\nu = 0 \ldots 2$ in the AdS base; $m,n = 1 \ldots 3$ on the 
sphere.  The flat tangent space indices are: $A,B = 0\ldots 5$, which 
parametrize six-dimensional (SO$(1,5)$) tangent vectors; $\alpha,\beta 
= 0 \ldots 2$, which index $\ads{3}$ (SO$(1,2)$) tangent vector indices; $a,b = 
1 \ldots 3$, indexing $S^3$ (SO$(3)$) tangent vectors.  The Kaluza-Klein 
gauge symmetry arising from the isometries of $S^3$ is $\so = \su 
\times \su$.  In our conventions, $I,J = 1 \ldots 6$ index $\so$, 
while $i,j = 1 \ldots 3$ index $\su$, as do $i',j'$. For Spin$(5)$, 
$\mathsf{i, j} = 1 \ldots 5$ labels the vector representation, while 
$r,s = 1 \ldots 4$ labels the spinors.  

We will not discuss the field
content of the tensor multiplets to which the minimal 
${\cal N} = 4b$ theory is coupled in detail. The only piece of
information that we need in the remainder is that tensor multiplets
contain two-form fields with anti self-dual three-form field strengths.
 
\subsection{Kaluza-Klein reduction revisited} 
 
Considerable work has been carried out on the topic of sphere 
compactifications (see the review~\cite{duffetal} and the recent 
works~\cite{spherecomp} for further references). The discussion below 
should serve as a review in a simplified setting. 
 
\paragraph{The metric: } 
A general compactification of six-dimensional gravity on a three dimensional 
compact space takes the form 
\begin{eqnarray} 
ds^2 &=& g_{\mu\nu} dx^{\mu} dx^{\nu} + g_{mn} Dx^m Dx^n \, , 
\label{metric} \\ 
Dx^m &=& dx^m - A^I_{\mu} K^m_I dx^{\mu} \, . 
\label{covder} 
\end{eqnarray} 
The Kaluza-Klein gauge fields $A^I_\mu$ are associated with the 
Killing vectors $K^m_I$ of the compact space.  (Note that the indices 
$I$ can be raised and lowered by the metric $\delta_{IJ}$.) 
 
We choose $g_{mn}$ to be the round metric on $S^3$.  Thus, we do not
include any scalars in our ansatz; as stated earlier, this is
motivated by the absence of scalar fields in the 3d Chern-Simons
supergravities with which we seek to make contact. Then there are six
Killing vectors arising from the $\so$ isometry group, and it is
manifest that the metric is invariant under $\so$ gauge
transformations:
\begin{eqnarray} 
\delta x^m & = & \eps^I K^m_I \, ,\\ 
\delta x^{\mu} & = & 0 \, ,\\ 
\delta A^I_{\mu} & = & \partial_{\mu} \eps^I + 
 f_{JK}{}^I A_{\mu}^J \eps^K \, . 
\end{eqnarray} 
Here $f_{JK}{}^I$ are the $\so$ structure constants, expressed in 
terms of the Killing vectors as 
\be 
f_{IJ}{}^K K_K^m = K^n_I \partial_n K^m_J - K^n_J \partial_n K^m_I . 
\ee 
The $\so$ gauge invariance of (\ref{metric}) follows from the 
transformations of $g_{mn}$ and $Dx^m$: 
\begin{eqnarray} 
\delta Dx^m &=& \eps^I \partial_n K^m_I Dx^n  \, ,\\ 
\delta g_{mn} &\equiv& \eps^I K_I^r \partial_r g_{mn} = 
-g_{rn} \eps^I \partial_m K^r_I - g_{mr} \eps^I \partial_n K_I^r \, . 
\end{eqnarray} 
Observe that $Dx^m$ transforms under a local gauge transformation 
in the same way as $dx^m$ under a global gauge transformation -- 
$D$ is like a covariant exterior derivative. 
 
\paragraph{The 3-form: } 
We must have a non-zero 3-form to satisfy the equations of motion. We 
will consider turning on just one of the five three-form fields 
$H^{\mathsf{i}}_{MNP}$. We require an $\so$ gauge invariant ansatz for this 
3-form field. Let 
\begin{equation} 
V(x^m) \epsilon_{mnr} dx^m \wedge dx^n \wedge dx^r \, , \qquad 
W(x^{\mu}) \epsilon_{\mu\nu\rho} dx^{\mu} \wedge dx^{\nu} \wedge 
dx^{\rho} 
\end{equation} 
be the volume forms on $S^3$ and on the non-compact factor in 
(\ref{metric}) respectively.  In terms of these forms, the six-dimensional 
equations of motion have an $\ads{3} \times S^3$ solution of the form 
(\ref{metric}) with vanishing Kaluza-Klein gauge fields and a 3-form background 
\begin{equation} 
H = {1 \over \ell} (W(x^{\mu}) \, \epsilon_{\mu\nu\rho} \, dx^{\mu} 
\wedge dx^{\nu} \wedge dx^{\rho} + V(x^m)\,  \epsilon_{mnr} \, dx^m 
\wedge dx^n \wedge dx^r),  
\end{equation}  
where $\ell$ is the radius of the $\sphere$. This cannot be quite 
right when the gauge fields are turned on, because it is not gauge 
invariant. A candidate gauge invariant generalization is 
\begin{equation} 
H = {1 \over \ell} (W(x^{\mu})\, \epsilon_{\mu\nu\rho} \, dx^{\mu} \wedge 
dx^{\nu} \wedge dx^{\rho} + V(x^m) \, \epsilon_{mnr}\, Dx^m \wedge 
Dx^n \wedge Dx^r) .  
\label{h2} 
\end{equation} 
Since the $S^3$ volume form is $\so$ invariant, 
$ \partial_m( K^m_I V(x^m))=0 $, 
(\ref{h2}) is gauge invariant. However, we should find a proposal for 
the 2-form potential $B_{MN}$, rather than the field strength $H$, 
which is only possible if $dH=0$.  The exterior derivative of 
(\ref{h2}) is computed using 
\begin{equation} 
dDx^m = - F^I K^m_I - A_{\mu}^I \partial_n K^m_I Dx^n \wedge dx^{\mu}, 
\end{equation} 
where $F^I = \frac{1}{2} F^I_{\mu\nu} dx^{\mu} \wedge dx^{\nu}$. We obtain 
\begin{equation} 
dH  =  -{3 \over \ell} V \epsilon_{mnr} K^m_I F^I \wedge Dx^n \wedge Dx^r, 
\label{tocancel} 
\end{equation} 
using the $\so$ invariance of the $S^3$ volume form  and the fact that 
one cannot anti-symmetrize over more than three indices. 
 
When the gauge field is flat (which is typically our interest) $dH=0$, 
as desired.  Nevertheless, it is worth seeking a more generally valid 
ansatz.  We wish to add a contribution to $H$ that cancels the term on 
the right hand side of (\ref{tocancel}).  To find this, it is helpful 
to consider the 2-form $\omega_I = V \, \epsilon_{mnr} \, K^m_I \,dx^n 
\wedge dx^r $ which appears as part of (\ref{tocancel}).  In terms of 
$\Omega$, the volume form on $S^3$, this 2-form can also be written as 
$\imath_{K_I} \Omega$. It is a standard fact that $ d \imath_{K_I} 
\Omega \, + \, \imath_{K_I}d \Omega = {\cal L}_{K_I} \Omega$.  Since 
the volume form is SO$(4)$ invariant, and annihilated by $d$, it 
follows that $\omega$ is closed.  Therefore, since we are on the three 
sphere there must be a globally well defined one-form $N_{Ir} dx^r$ such 
that $d(N_{Ir} dx^r) = \omega$.  Assembling these facts, a candidate 
Kaluza-Klein ansatz for a closed 3-form is 
\begin{equation} 
H_{KK} = H + {3 \over \ell} F^I \wedge N_{Ir} Dx^r 
\label{hkk} 
\end{equation} 
The 1-forms $N_{Ir} \, dx^r$ for $S^3$ are related to the Killing 
one-forms and are derived explicitly in Appendix~\ref{sphereapp}. The 
choice of $N_{Ir}$ given there satisfy the relation 
\begin{equation} 
K^m_J \partial_m N_{Ir} + N_{It} \partial_r K^t_J = 
f_{JI}{}^K N_{Kr}. 
\label{stuff} 
\end{equation} 
Using this relation it can be checked that $H_{KK}$ is still gauge 
invariant, and that 
\begin{equation} 
d(F^I N_{Ir} Dx^r) =  
V \epsilon_{mnr} K^m_I F^I \wedge Dx^n \wedge Dx^r. 
\end{equation} 
Combining this with (\ref{tocancel}) shows that $H_{KK}$ is a closed 
form, as desired.  Thus, we have a consistent $\so$ invariant ansatz 
for Kaluza-Klein reduction of six dimensional gravity on a sphere, 
with gauge field VEVs. 

Notice that the three-form $H_{KK}$ is not  self-dual. Therefore,
this ansatz cannot be given for the minimal 
${\cal N} = 4b$ theory, but we need at least one tensor
multiplet as well. The self-dual part of $H_{KK}$ then lives
in the gravity multiplet, the anti self-dual part lives in the
tensor multiplet. Together,  one self-dual and one anti
self-dual tensor combine into an unconstrained two-form 
field. We can think of such a two-form field as originating
in either the NS or RR two-form in type IIB string theory in
ten dimensions. In particular, for the equations of motion we
can use the equations of motion of string theory, rather than 
the more complicated ones of ${\cal N} = 4b$ supergravity. 
 
\paragraph{Equations of motion: }  Using the results collected 
in~\cite{duffetal} and the above remarks, it is now a straightforward,
if lengthy, exercise to compute the six-dimensional equations of
motion for our Kaluza-Klein ansatz.  As in~\cite{duffetal}, it is
easier to work out the equations of motion using the vielbein
formalism. It is convenient to display the the $\so = \su \times \su$
gauge symmetry inherited from isometries of the sphere explicitly by
picking a basis of Killing vectors such that the left ($F_L^i$ ,
i=1,2,3) and right ($F_R^{i'}$ , i' = 1,2,3) $\su$ field strengths
are:
\begin{eqnarray} 
F^I_{\alpha\beta} &=& F_{L\alpha\beta}^I  ~~~~~~  I=1,2,3\\ 
                  &=& F_{R\alpha\beta}^{I-3} ~~~~ I=4,5,6 
\end{eqnarray} 
Such a basis is explicitly constructed in Appendix~\ref{sphereapp}. In 
simplifying the equations of motion, the following identities are 
useful. First, one can show that 
\be 
K^m_I g_{mn} K^n_J + {1 \over \ell^2} N_{Im} g^{mn} N_{Jn} = {\ell^2 
\over 2} \delta_{IJ} . 
\ee 
Second, there is a simple map from SO$(4)$ to itself, 
that acts as $+1$ on SU$(2)_L$ and as $-1$ on SU$(2)_R$,
which we will denote by $A_I^J$. In other words, 
it sends $K^m_I$ to $A_I^J K^m_J$. Then we have 
\be 
g_{mn} K^n_I = {1 \over \ell} A_I^J N_{Jm}  . 
\ee 
Then, if we take the metric $g_{\mu\nu}$ and the Kaluza-Klein gauge 
fields $A_\mu^I$ to only depend on the coordinates $x^\mu$ of the 
three-dimensional non-compact space, the ansatz will satisfy all the 
equations of motion of the  
six-dimensional theory if the metric and gauge field satisfy the 
following three-dimensional equations : 
\begin{eqnarray} 
R_{\alpha\beta} + {2 \over \ell^2} \delta_{\alpha\beta} - 
\frac{1}{2} \delta_{IJ} F^I_{\alpha\gamma} F^J{}_{\beta}{}^{\gamma} 
& = & 0 \label{eom1} \, ,\\ 
D \ast F^{(L)} + F^{(L)} 
+g(D \ast F^{(R)} - F^{(R)}) g^{-1} & = & 0 \label{eom3} \, ,\\ 
{\rm tr}(F^{(L)}_{\beta\gamma}g \partial_m g^{-1})
{\rm tr}( F^{(R)\beta\gamma}g^{-1}\partial_n g) & = & 0 \, 
\label{eom4}  \\
{\rm tr}(F^{(L)} - g F^{(R)} g^{-1})^2 & = & 0 \, . \label{eom5}
\end{eqnarray} 
Here, we used a group element $g\in$SU$(2)$ to parameterize
the $S^3$, and SU$(2)_{L,R}$ correspond to the left and right
actions on the three-sphere. The last equation of motion 
(\ref{eom5}) has its origin in the dilaton equation of motion.
It is clear that the equations of motion
are gauge invariant, and that
any solution to three dimensional cosmological gravity with flat gauge 
fields solves these equations.  These are the solutions of the bosonic 
part of the ${\rm SU(1,1|2)} \times {\rm SU(1,1|2)}$ Chern-Simons 
supergravity, and include the conical defects: 
\begin{eqnarray} 
ds^2 &=& -\left( {r^2 \over \ell^2} - M_3 \right) \, dt^2 + 
\left( {r^2 \over \ell^2} - M_3 \right)^{-1} \, dr^2 + r^2 \, d\phi^2 
\ , \\ 
F_L^i &=& 0 ~~~~~;~~~~~ F_R^{i'} = 0 \ . 
\end{eqnarray} 
However, although (\ref{eom1})--(\ref{eom5}) allow $F^{(L)} =
F^{(R)}=0$ they do not obviously {\it imply} this.  
If they did, we would have found a consistent truncation of the
six-dimensional theory to three-dimensional Chern-Simons theory.
Notice that the first two equations of motion (\ref{eom1}) and 
(\ref{eom3}) can naturally be obtained from
a three-dimensional theory
consisting of the Einstein-Hilbert term, a Yang-Mills term and a
Chern-Simons term. The other two equations (\ref{eom4}) and (\ref{eom5})
do not have such a clear interpretation.
It has been shown in~\cite{spherecomp} that consistent
Kaluza-Klein reductions with general $\so$ gauge fields can be
achieved by also turning on scalar fields that parameterize the shape
of the compact manifold.
 
Thus, although the $SU(1,1|2) \times SU(1,1|2)$ Chern-Simons 
supergravity in 3 dimensions has the symmetries of the six-dimensional 
theory reduced on a sphere, our ansatz 
does not produce this theory.  The Chern-Simons formulation of 
$\ads{3}$ supergravity has been an important tool in investigations of 
the AdS/CFT correspondence (see, e.g.,~\cite{janads,david:point,david:group} 
amongst many other references).  While many of these works relied 
primarily on symmetries, it remains desirable to explain precisely how 
and whether the six-dimensional, ${\cal N} = 4b$ gravity reduces to 
the three-dimensional $SU(1,1|2) \times SU(1,1|2)$ theory.  Once we 
include scalars, we can obtain consistent truncations to a 
three-dimensional action. Although these theories have more than just 
a Chern-Simons term, at low energies they can be approximated by a 
Chern-Simons theory -- the $F^2$ terms in the action can be 
ignored at low energy. A more precise argument is given in ~\cite{csym},
where is it shown that wavefunctions in the Yang-Mills Chern-Simons
theory can be decomposed in a natural way in a Yang-Mills piece
and a Chern-Simons piece.

We should also comment on the relation between our Kaluza-Klein ansatz
and the results in section~7 of ~\cite{deger:6d}, where a Chern-Simons
like structure is found for the field equations for a certain set of
gauge fields. The computation in ~\cite{deger:6d} differs from ours in
several ways. First of all, the gauge fields appearing in the three
form and the metric of their Kaluza-Klein ansatz are different. Thus,
the dimensionally reduced theory has two different ``gauge fields,''
but only one gauge invariance.  Secondly, they only consider the
self-dual three-form, whereas our KK ansatz contains both a self-dual
and an anti-self-dual three-forms. In particular, equation (152)
in~\cite{deger:6d} depends explicitly on the gauge fields, and is a
consequence of the self-duality equation for the three-form.  In our
case we do not impose such a self-duality relation, and as a
consequence, we do not find a field equation of the form (152). The
field equation (\ref{eom4}) is not obtained in~\cite{deger:6d},
because they only consider the linearized system.

The results of \cite{deger:6d} were extended in \cite{stefan} where
not only quadratic but also cubic couplings in the six-dimensional
theory were considered. It was found that, to that order, there
exists a gauge field whose field equation becomes the Chern-Simons
field equation and that massive fields can be consistently put to
zero. The gauge field in question is a linear combination of the
gauge fields appearing in the metric and in a self-dual two-form. 
If we were to insist that our three-form is self-dual, we would
also find the Chern-Simons field equation, and in this sense the
results agree with each other.

\paragraph{Summary: } We have found an $\so$ invariant 
Kaluza-Klein ansatz for the $S^3$ compactification of six dimensional 
supergravity, involving just the KK gauge fields and no scalars.  Upon 
dimensional reduction, however, we do not find equations of motion 
that could arise from a three dimensional effective action. In any 
case, if $F=0$, our ansatz for the metric and $H_{KK}$ provide 
solutions to the 6d equations of motion.  The effective 3-dimensional 
equations are solved by any solution to three dimensional 
cosmological gravity with a flat gauge field.  This 
spectrum of solutions includes the supersymmetric conical defects we 
are interested in.  Below we will show how the gauge fields are chosen 
to make the solutions supersymmetric. 
 
\subsection{Supersymmetry} 
 
Having found an appropriate Kaluza-Klein ansatz, we investigate the
supersymmetry of the solutions incorporating conical defects.  By
examining the Killing spinor equations, with a flat KK gauge field, we
recognize the effective 3d equations as the Killing spinor equations
of the $SU(1,1|2) \times SU(1,1|2)$ Chern-Simons supergravity.  This
allows us to use the work of~\cite{Izq:2+1,david:point} to choose a
Wilson line for which the 3d conical defects lift to supersymmetric
solutions of the six-dimensional theory.

\subsubsection{6d Killing spinor equations} 
 
First, the 10d IIB supergravity has 32 supersymmetries. Half of them
are broken by the reduction on K3, so we are left with 16
supersymmetries in six dimensions. The resulting theory is the N=4b
supergravity in six dimensions. As long as we consider flat gauge
fields, the three-form is self-dual, and we can ignore the tensor
multiplets.  N=4b supergravity is a chiral theory, with four chiral,
symplectic-Majorana supercharges (labeled by $r=1,\ldots 4$), each
having four real components.  Following Romans~\cite{romans}, the
${\cal N}=4b$ algebra can be viewed as an extension of an $N=2$
algebra. The $N=2$ algebra is generated by a doublet of chiral
spinorial charges, and it has an ${\rm USp}(2)= \su$ R-symmetry. The
charges are doublets under the $\su$. The ${\cal N}=4b$ algebra can be
viewed as an extension of $N=2$ to $N=4$, where one takes two copies
of the $N=2$ charges of the same chirality.  The resulting algebra has
an ${\rm USp}(4)= {\rm Spin}(5)$ R-symmetry, and the four supersymmetry
parameters $\epsilon_r$ transform in the fundamental representation of
${\rm Spin}(5)$.
 
${\rm Spin}(5)$ is represented by the $4\times 4$ Gamma matrices 
$\Gamma^{\mathsf{i}}$:  
\begin{equation} 
      \{ \Gamma^{\mathsf{k}} , \Gamma^{\mathsf{l}} \} = 
\delta^{\mathsf{kl}} \ , \  \mathsf{k,l}=1,\ldots  , 
      5. 
\end{equation} 
$\Gamma^5$ has two $+1$ eigenvalues, and two $-1$ eigenvalues. Hence, by 
taking suitable linear combinations of the supersymmetry parameters 
$\epsilon_r$, we can organize things so that 
\begin{equation} 
   ( \Gamma^5 )_{rs} \epsilon_s = \left\{ 
   \begin{array}{l} +\epsilon_r \ , \ {\rm for}\ r=1,3 \\ 
                    -\epsilon_r \ , \ {\rm for}\ r=2,4 \ . 
                    \end{array} \right. 
\label{gamma5} 
\end{equation} 
The 6d Killing spinor equation is 
\begin{equation} 
D_M \epsilon_r - \frac{1}{4} H^{\mathsf{k}}_{MNP} \Gamma^{NP} 
(\Gamma^{\mathsf{k}} )_{rs} \epsilon_s 
=0\ . 
\label{ks} 
\end{equation} 
In our solutions only one of the five three form fields is turned on, 
and by U-duality, we can choose $H^{\mathsf{k}}_{MNP} \sim 
\delta^{\mathsf{k}5}$.  
When the field strengths $F^I$ vanish, the gauge invariant definition
of $H$ in (\ref{hkk}) reduces to (\ref{h2}).  For the $M=\mu$
components of the Killing spinor equation, the relevant components of
the three form field are thus:
\begin{equation} 
H^5_{\alpha\beta\gamma} = \ell^{-1} \epsilon_{\alpha\beta\gamma} 
 \ ; \ 
H^5_{abc} = \ell^{-1} \epsilon_{abc} \ ; \ 
  H^5_{\mu a b} = -\ell^{-1}K^m_I A^I_{\mu}e_m^c \epsilon_{abc} \ . 
\end{equation} 
$\Gamma^5$ can be dropped from the Killing spinor equation with 
the help of (\ref{gamma5}). \noindent For the purposes of 
Kaluza-Klein reduction, we also decompose the SO$(1,5)$ gamma 
matrices $\Gamma^A$ as direct products of SO$(3)$ and SO$(1,2)$ 
matrices ($\gamma^a$ and $\gamma^{\alpha}$) as follows: 
\begin{eqnarray} 
\Gamma^\alpha &=& \sigma^1 \otimes 1\otimes \gamma^\alpha \ ; \ 
\Gamma^a = \sigma^2 \otimes \gamma^a \otimes 1  \, , \\ \gamma^0 
&=& -i\sigma_2 \ ; \ \gamma^1 = \sigma_1 \ ; \ \gamma^2 = 
 \sigma_3 ~~~~~;~~~~~ 
\gamma^a = \sigma^a \ , \ a=1,2,3 
\end{eqnarray} 
Then, for example, we get 
$ \Gamma^{\alpha\beta} =  
1\otimes 1\otimes \epsilon^{\alpha\beta\delta} \gamma_\delta; 
$ 
and 
$ 
\Gamma^{ab} =   
1\otimes i \epsilon^{abc} \gamma_c \otimes 1. 
$ 
 
Note that the 6d gamma matrices are 8x8, but the chiral 
spinors in 6d have 4 components. Chiral spinors $\Psi^{(\pm)}$ satisfy 
\begin{equation} 
    \Psi^{(\pm)} = \frac{1}{2} (1\pm \Gamma^7 ) \Psi 
\end{equation} 
where $\Gamma^7 = \Gamma^0 \Gamma^1 \cdots \Gamma^5 = \sigma_3 
\otimes 1 \otimes 1 $.  We let the ${\cal N}=4b$ spinors be of 
positive chirality ($\Psi^{(+)}$). Then, in the Killing spinor 
equation (\ref{ks}), all the supersymmetry parameters $\epsilon_r$ 
are of the form 
\begin{equation} 
         \epsilon_r = \left( \begin{array}{c} \varepsilon_r \\ 0 
         \end{array} \right) \ , 
\label{epsform} 
\end{equation} 
where $\varepsilon_r$ is a doublet of two-component spinors. 
We can additionally impose a symplectic Majorana condition on 
these spinors~\cite{romans}.  It then follows, as is shown in detail in 
Appendix~\ref{sympapp}, that $\varepsilon_r$ can be written as 
an $\su$ doublet of complex conjugate two-component spinors: 
\be 
  \varepsilon_r = \left( \begin{array}{l} 
  \varepsilon^{(2)}_r \\ \varepsilon^{(2)*}_r \end{array} \right) \ . 
\label{su2doublet} 
\ee 
 
Consider first the $M=m$ internal component of the Killing spinor 
equation: 
\begin{equation} 
\left( D_{m} \mp \frac{1}{4}H^5_{m NP} \Gamma^{NP} \right) 
\epsilon_r = 0. 
\end{equation} 
The upper signs and lower signs ($-$ and $+$) correspond to $r=1,3$ and 
$r=2,4$ respectively. This split will relate to the $SU(2)_L$ and 
$SU(2)_R$ sectors. We assume that the Killing spinor is in a zero mode on 
the sphere, in accord with our Kaluza-Klein approach. That is, 
$\epsilon_r$ is independent of the sphere coordinates, so that 
\begin{equation} \label{cov} 
D_m \epsilon_r = (\p_m + \frac{1}{4} \hat{\omega}_m^{\ AB}\Gamma_{AB}) 
\epsilon_r = \frac{1}{4} \hat{\omega}_m^{\ AB}\Gamma_{AB} \epsilon_r = 
\frac{i}{4} \epsilon_{abc} \omega_m^{\ ab} 1\otimes \gamma^c \otimes 1 
\ \epsilon_r \ . 
\end{equation} 
The three-form contribution is 
\begin{eqnarray} 
\mp \frac{1}{4} H^5_{mNP} \Gamma^{NP} 
 &=& \mp \frac{1}{4} H^5_{mnp} \Gamma^{np} \nonumber \\ 
\mbox{} &=& \mp \frac{i}{4\ell} e^a_m \epsilon_{abc} 
\epsilon^{bcd} 1\otimes \sigma_d \otimes 1 \nonumber \\ 
\mbox{} &=& \mp \frac{i}{2\ell} e^a_m 1\otimes \sigma_a \otimes 1 \ . 
\end{eqnarray} 
Thus, the internal Killing spinor equation is  
\begin{equation} 
\label{ksp} 
 \frac{i}{4} (\epsilon_{abc} \omega_m^{bc} \mp \frac{2}{\ell} e_{ma} ) 
 (1\otimes \sigma^a\otimes 1) \epsilon_r = 0 \ . 
\end{equation} 
Now, $g_{mn}$ is by assumption the metric of a round three-sphere. We 
can show by explicit calculation, using the bases for $S^3$ in 
appendix B, that 
\begin{equation} \label{idl} 
\epsilon^{abc} \omega_{bc} = \frac{2}{\ell} e^a 
\end{equation} 
when we use the basis $e_a = {2 \over \ell} L_a$, and 
\begin{equation} \label{idr} 
\epsilon^{abc} \omega_{bc} = - \frac{2}{\ell} e^a 
\end{equation} 
when we use the basis $e_a = - {2 \over \ell} R_a$. Thus, the internal 
Killing spinor equation can be trivially satisfied. This is as we might 
have expected; since our Kaluza-Klein ansatz leaves the form of the 
metric $g_{mn}$ fixed, the internal Killing spinor equation is always 
the same, and we know it is satisfied in the AdS$_3 \times S^3$ 
vacuum. 
 
Consider now the $M=\mu$ component of the 6d Killing spinor equation: 
\begin{equation} 
\left( D_{\mu} \mp \frac{1}{4}H^5_{\mu NP} \Gamma^{NP} \right) 
\epsilon_r = \left( D_{\mu} \mp \frac{1}{2\ell} 
e_{\mu\alpha}(1\otimes 1 \otimes \gamma^{\alpha})  \pm 
\frac{i}{2\ell} A^I_{\mu} K^m_I (1\otimes \sigma_m \otimes 1) 
\right) \epsilon_r =0\ . \label{Kspmu} 
\end{equation} 
As before, the upper signs and lower signs correspond to $r=1,3$ 
and $r=2,4$ respectively. The gauge covariant derivative is~\cite{romans} 
\begin{eqnarray} 
D_{\mu}\epsilon_r &=& \partial_{\mu} \epsilon_r +\frac{1}{4} 
\hat{\omega}^{AB}_{\mu} \Gamma_{AB} \epsilon \label{cder} \\ 
\hat{\omega}^{AB}_{\mu} \Gamma_{AB} &=& \omega^{\alpha \beta}_{\mu} 
\Gamma_{\alpha \beta} -  A^I_{\mu} \nabla_a K_{Ib} 
\Gamma^{ab} \, , 
\label{omehat} 
\end{eqnarray} 
Using the definition of the gamma matrices, the last term of 
(\ref{omehat}) becomes 
\begin{eqnarray} 
A^I_{\mu} \nabla_a K_{Ib} \Gamma^{ab} &=& A^I \nabla_a K_{Ib} i\epsilon^{abc} 
(1\otimes \sigma_c \otimes 1) \, , \\ 
 \nabla_a K_{Ib} &=& \frac{1}{\ell^2}\epsilon_{abc} N^c_I \, . 
\end{eqnarray} 
where we have used the relation between the Lorentz covariant 
derivative of $K$ and the components of a one-form $N$ (see 
Appendix~\ref{sphereapp}). 
Folding these facts into the last term in (\ref{cder}) yields 
the Killing spinor equation (\ref{Kspmu}) as: 
\begin{equation} 
 \left( 
 \partial_\mu 1\otimes 1 +\frac{1}{4} \epsilon_{\alpha \beta 
  \delta }\omega^{\alpha \beta}_{\mu} 1\otimes \gamma^{\delta} 
  +\frac{i}{2\ell} A^{I}_{\mu} (-\frac{1}{\ell}N^c_I \mp K^c_I) \sigma_c \otimes 1 
  \pm \frac{1}{2\ell} e_{\mu\alpha} 1\otimes \gamma^{\alpha} \right) 
  \varepsilon_r = 0 \ , 
\end{equation} 
where we used (\ref{epsform}) for the chiral spinors.   Now, 
according to Appendix~\ref{sphereapp}, the combinations 
$\ell^{-1}N^I_c \pm K^I_c$ are projectors to the left and right 
$SU(2)$ sectors, 
\begin{eqnarray} 
 L^I_c &=& - \frac{1}{\ell}N^I_c + K^I_c = \left\{ \begin{array}{l} 
   \ell \delta^I_c \ \ \ \ \, {\rm for} ~ \ I=1,2,3 \\ 0 \ \ \ \ \ \  \
{\rm for} ~ \
I=4,5,6 \, , 
   \end{array} \right.  
   \\ 
 R^I_c &=& - \frac{1}{\ell}N^I_c - K^I_c = \left\{ \begin{array}{l} 
   0 \ \ \ \ \  \ \ {\rm for} ~ \ I=1,2,3 \\ \ell \delta^{I-3}_c \ \ {\rm for} ~ \ 
   I=4,5,6 \, .  \end{array}\right. 
\end{eqnarray} 
Then, the two Killing spinor equations labeled by $r=1,3$ ($r=2,4$) give the 
$\su_L$ ($\su_R$) sector equations: 
\begin{equation} 
\left( 
 \partial_\mu 1\otimes 1 +\frac{1}{4} \epsilon_{\alpha \beta 
  \delta }\omega^{\alpha \beta}_{\mu} 1\otimes \gamma^{\delta} 
  +\frac{i}{2} A^{c}_{\mu} \sigma_c \otimes 1 
  - \frac{1}{2\ell} e_{\mu\alpha}1\otimes \gamma^{\alpha} \right) 
  \varepsilon_r = 0 \, , \label{su2l} 
\end{equation}
for $r=1,3$ and 
\begin{equation}
\left( 
 \partial_\mu 1\otimes 1 +\frac{1}{4} \epsilon_{\alpha \beta 
  \delta }\omega^{\alpha \beta}_{\mu} 1\otimes \gamma^{\delta} 
  +\frac{i}{2} A'^{c}_{\mu} \sigma_c \otimes 1 
  + \frac{1}{2\ell} e_{\mu\alpha}1\otimes \gamma^{\alpha} \right) 
  \varepsilon'_r = 0 \,  . \label{su2r} 
\end{equation} 
for $r=2,4$. Because of the doublet structure (\ref{su2doublet}), each
spinor $\varepsilon_r$ has four real degrees of freedom. Since we have
two equations in the $\su_L$ sector and two in the $\su_R$ sector, in
total we have 8+8=16 supersymmetry parameters, in agreement with the
16 supersymmetries of the 6d theory.  From the three dimensional point
of view of the $\ads{3}$ base of our fibered compactification, this is
the $N=(4,4)$ supersymmetry, since $N$ counts the number of
supercharges, which in 3d are real two-component spinors.  Below, we
will use the results of~\cite{Izq:2+1,david:point} to choose a
Kaluza-Klein Wilson line for our 6d solutions that makes them
supersymmetric.
 
\subsubsection{$SU(1,1|2)\times SU(1,1|2)$ supergravity} 
 
We now compare the three-dimensional spinor equations 
(\ref{su2l},\ref{su2r}) to the Killing spinor equations for the 
three-dimensional $SU(1,1|2)\times SU(1,1|2)$ supergravity. The latter 
is described by the action \cite{NT,david:group} 
\bea  
  S &=& \frac{1}{16\pi G} \int d^3x [eR +\frac{2}{\ell^2}e 
  \nonumber \\ 
 +i\varepsilon^{\mu\nu\rho}\bar{\psi}_{\mu r} {\cal D}_{\nu} \psi_{\rho r} &-& 
 \ell \varepsilon^{\mu\nu\rho}~{\rm Tr} (A_{\mu}\pat_{\nu} A_{\rho} 
  +\frac{2}{3}A_{\mu}A_{\nu}A_{\rho}) 
  \nonumber \\ 
 +i\varepsilon^{\mu\nu\rho}\bar{\psi}'_{\mu r} {\cal D}'_{\nu} \psi'_{\rho r} &+& 
 \ell \varepsilon^{\mu\nu\rho}~{\rm Tr} (A'_{\mu}\pat_{\nu} A'_{\rho} 
  +\frac{2}{3}A'_{\mu}A'_{\nu}A'_{\rho})] 
  \ , \label{davidact} 
\eea 
where $e^{\alpha}_\mu$ is the dreibein, 
$A_\mu$ and $A'_\mu$ are the $SU(2)_L$ and $SU(2)_R$ gauge 
fields 
\be 
     A_\mu = A^a_{\mu} \frac{i\sigma_a}{2} \ \ , \ \ A'_\mu = 
     A'^{a}_\mu \frac{i\sigma_{a}}{2} \ , 
\ee 
and $\psi_{\mu r}$  ($\psi'_{\mu r}$) with $r=1,2$ are the 
$SU(2)_L$ ($SU(2)_R$) doublet two-component spinors of Appendix A. 
The covariant derivatives are 
\bea 
\cal{D}_{\mu} &=& \pat_{\mu} + \frac{1}{4} \omega_{\mu \alpha \beta} 
 \gamma^{\alpha \beta} 
 + A_\mu -\frac{1}{2\ell}e_{\mu \alpha} \gamma^\alpha \\ 
\cal{D}'_{\mu} &=& \pat_{\mu} + \frac{1}{4} \omega_{\mu\alpha\beta} 
  \gamma^{\alpha \beta} 
 + A'_\mu + \frac{1}{2\ell}e_{\mu\alpha} \gamma^\alpha \ . 
\eea 
Recall that $\gamma^{\alpha \beta} 
= (1/2)[\gamma_\alpha ,\gamma_\beta]=\varepsilon^{\alpha\beta\delta} 
\gamma_\delta$. 
Recall that in three spacetime dimensions there are two inequivalent
two-dimensional irreducible representations for the $\gamma$-matrices
($\gamma$ and $-\gamma$) (see~\cite{AchuTown,coushen}).  The two
sectors in the action (\ref{davidact}) are related to the two inequivalent
representations. Therefore, the two covariant derivatives ${\cal D}$ differ by a
minus sign in the $\gamma$-matrices.
 
The supersymmetry transformation of the spinors gives the 
Killing spinor equations 
\be 
    \delta \psi_{\mu r} = {\cal D}_{\mu} \epsilon_r = 0 \ 
\ ; \  \ \delta \psi'_{\mu r} = {\cal D}'_{\mu} \epsilon'_r = 0 \ . 
\label{epsilons} 
\ee  
One can readily see that the equations (\ref{epsilons}) are identical to 
(\ref{su2l},\ref{su2r}). 
The solution of these equations for the point 
particle spacetimes was already considered in the context of the 
$SU(1,1|2)\times SU(1,1|2)$ supergravity in 
\cite{david:point}. However, \cite{david:point} presents a rather 
brief discussion of the actual embedding of the solutions of 
\cite{Izq:2+1}, leaving out many issues that are relevant to us.  We 
therefore give a complete discussion of the solution of 
(\ref{su2l},\ref{su2r}), using the results of \cite{Izq:2+1},
in the next two subsections. 
 
\subsubsection{Conical defects as BPS solutions in $(2,0)$ supergravity} 
 
We have reduced the problem of finding the Killing spinors in 6d 
supergravity to solving (\ref{su2l},\ref{su2r}) in 2+1 
dimensions. Then the task has been made much easier, since a related 
problem has already been solved in \cite{Izq:2+1}. We only need a 
minor generalization of the solutions of \cite{Izq:2+1} to construct 
solutions for our equations. In this and the following section we will 
show in detail how to do the embedding. In particular, we are 
interested in keeping track of the number of supersymmetries that are 
preserved as the conical deficit parameter increases from $0$ to its
extreme value.

Extended $\ads{3}$ supergravity theories were first constructed 
based on the $Osp(p|2,R)\otimes OSp(q|2,R)$ supergroups 
\cite{AchuTown}, and are referred to as $(p,q)$ supergravities.  The 
number of supercharges is $N=p+q$, and each of them is a two-component 
real spinor.  The action also contains $O(p)\times O(q)$ gauge 
fields. Izquierdo and Townsend~\cite{Izq:2+1} embedded the 3d conical defects
into (2,0) supergravity and investigated 
their supersymmetry. In \cite{Izq:2+1}, the two-component real spinors 
have been combined into a single complex spinor, so the $O(2)$ gauge 
group has been interpreted as a $U(1)$. Then there is a single complex 
vector-spinor gravitino field, with a supersymmetry transformation 
parameterized by a single complex two-component spinor parameter.  The 
corresponding Killing spinor equation is 
\begin{equation} 
   \CD_{\mu} \epsilon =0 \label{itks} 
\end{equation} 
with the covariant derivative \footnote{In converting from the 
$(+--)$ signature of \cite{Izq:2+1} to our $(-++)$ signature, we have 
replaced $\gamma^{\alpha}$ by $-i\gamma^\alpha$. 
Note that \cite{Izq:2+1} uses a different notation, with 
$ 
            \frac{1}{2\ell} = m \ . 
$ 
} 
\begin{equation} 
  \CD_{\mu} = \pat_{\mu} + \frac{1}{4}\epsilon_{\alpha \beta \delta} 
  \omega_{\mu}^{~~\alpha \beta }\gamma^{\delta}+\frac{i}{\ell} A_{\mu} 
  -\frac{1}{2\ell} e_{\mu\alpha} \gamma^{\alpha} \ . 
\end{equation}

Izquierdo and Townsend find {\em two} Killing spinors (out of the maximum 
of four, counting the real degrees of freedom) for conical defects with 
Wilson lines. The three-dimensional metric we are interested in is 
(\ref{ppmetric}) with $M_3 = -\gamma^2$. The $U(1)$ gauge potential 
producing to the Wilson line is  
\begin{equation} \label{jj8}
   A = -\frac{\ell}{2} (\gamma + n) d\phi, 
\end{equation} 
where $n$ is an integer related to the periodicity of the Killing 
spinors. If $\gamma =-n$, the gauge field is zero. If, in addition,  
$\gamma =\pm 1$ we recover a global adS$_3$ metric. The case $n=0$, 
$0<|\gamma |<1$ corresponds to the point mass spacetimes in which we
are interested. These have charge 
\begin{equation}
      Q  = \frac{1}{2\pi \ell} \oint A = 
      -\frac{\gamma}{2}  \, ,
\end{equation}
so that $M=-4Q^2$. The deficit angle is $\Delta \phi = 2\pi (1-|\gamma 
|)$, as we saw at the beginning of this section.  The origin $r=0$ is 
a conical singularity and is excised from the spacetime. 
 
The Killing spinor solution is~\cite{Izq:2+1} 
\bea 
   \epsilon &=& e^{in\phi /2 + i\gamma t/2\ell}~[k_-\sqrt{f+\gamma} 
   -k_+\sqrt{f-\gamma}]  \label{itspinor} \\ 
   \mbox{} &\times& \left\{ \left[ 
   1-\frac{1}{f} (i\gamma \gamma^0+\sqrt{f^2-\gamma^2}~\gamma^1)\right] 
   -ib_2\gamma^2\left[1+\frac{1}{f}(i\gamma \gamma^0 
+\sqrt{f^2-\gamma^2}~\gamma^1)  
   \right] \right\} \zeta_0, \nonumber 
\eea 
where $k_{\pm}$ are arbitrary constants, 
\be 
  b_2 = \frac{k_+\sqrt{f+\gamma}+k_-\sqrt{f-\gamma}} 
  {k_-\sqrt{f+\gamma}-k_+\sqrt{f-\gamma}} \ , 
\ee 
and $\zeta_0$ is a constant spinor. It satisfies a 
projection condition $P\zeta_0=\zeta_0$ with the projection matrix 
\be 
  P = \frac{-1}{(k^2_++k^2_-)}~\left[ i(k^2_- -k^2_+)\gamma^0 
  -2k_+k_-~\gamma^1 \right] \ . 
\ee 
For fixed $k_{\pm}$, the projection removes two of the four real
spinor degrees of freedom, so the space of Killing spinors $\epsilon$
has two real dimensions. Note that Izquierdo and Townsend find Killing
spinors for arbitrary $\gamma, n$. Apparently this leads to BPS
solutions  of arbitrarily negative mass. We will comment briefly on their
meaning in section~\ref{sec:spin}.
 
The Killing spinors may be singular at $r=0$. Near the origin, 
$\epsilon$ behaves as 
\be 
    \epsilon \sim r^{\frac{\sigma}{2}}e^{in\phi /2} \epsilon_0 
\label{limit1}  
\ee  
where $\epsilon_0$ is some constant spinor and 
$\sigma$ depends on $\gamma, n$. If $\sigma$ is a positive 
integer, $\epsilon$ will be regular at the origin. If $\sigma =0$, 
the spinor will be regular if $|n|=1$, but otherwise it is 
singular. For $\sigma <0$ the spinor is singular. 
 
When $n=0$, $0<|\gamma | <1$, corresponding to the conical defects,
$\sigma=0$ in (\ref{limit1}) but $n=0$, the Killing spinors are
periodic, and, since we are working in a polar frame, singular at the
origin.  However, the origin is in any case a singular point, and removed
from the spacetime. That is to say, the
spacetime has noncontractible loops so $Q\neq 0$ is possible. There
are then two Killing spinors.
 
Let us consider the case of global AdS$_3$ in greater detail. AdS$_3$ in 
global coordinates with zero gauge fields is obtained when
$\gamma = -n = \pm 1$. In this case, the origin becomes regular. The 
corresponding Killing spinors have $\sigma =0$ and are regular at the 
origin, as required. They are antiperiodic in $\phi$, as expected 
since the space is now contractible. We get two Killing spinors with 
$\gamma = -n = 1$, and two with $\gamma = -n = -1$. Since both these 
choices give the AdS$_3$ geometry, we see it has four Killing 
spinors, that is, it preserves the full supersymmetry of $(2,0)$ supergravity.
 
What is the relation between global $\ads{3}$ and the conical defects
with Wilson lines?  There are two limits of the point particles. The
limit $n = \gamma =0$ corresponds to the $M=J=Q=0$ black hole vacuum,
and it has two Killing spinors. One can move away from this limit in
either the $\gamma >0$ direction or the $\gamma <0$ direction. The
limit $\gamma = \pm 1$, $n=0$ corresponds to AdS$_3$ with non-zero
gauge fields of charge $Q = \mp {1 \over 2}$. Now note that 
the integer $n$ can be changed by a large gauge
transformation~\cite{Izq:2+1} (from the six-dimensional point of view,
this corresponds to a coordinate transformation on $S^3$; see section
\ref{sec:spin} for details). In section~4, we will see that such
large gauge transformations correspond to a spectral flow in the
boundary CFT.  For $\gamma = \pm 1$, we can make a gauge
transformation to make $n = \mp 1$; this turns the periodic spinors
associated with the point particle geometries into the antiperiodic
spinors associated with AdS$_3$. Again, AdS$_3$ has twice as many
supersymmetries, because there are two ways to reach the AdS$_3$
limit.
 
\subsubsection{Embedding into 6d $N=4b$ supergravity} 
 
It is quite simple to promote Izquierdo's and Townsend's solutions 
for $(2,0)$ Killing spinors to solutions of the Killing spinor equations 
(\ref{su2l},\ref{su2r}). To relate the Killing spinor equation 
(\ref{itks}) to (\ref{su2l}), we replace the $U(1)$ gauge 
potential by a $SU(2)_L$ gauge potential, 
\be 
   \frac{1}{\ell} A^{U(1)}_\mu  \rightarrow \frac{1}{2} 
   A^{SU(2),c}_\mu \sigma_c \ , 
\ee 
and the spinor by the $SU(2)_L$ doublet of spinors, 
\be 
  \epsilon \rightarrow \epsilon_r = 
  \left( \begin{array}{c} \varepsilon_r \\ \varepsilon^*_r 
  \end{array} \right) \ . 
\ee 
Recall that the label $r=1,3$ is needed, since (\ref{su2l}) contains 
two identical Killing spinor equations. 
The $U(1)$ Wilson line is embedded into the $SU(2)$ by 
\be \label{jj9}
   \frac{1}{\ell} A^{U(1)}_\phi = -\frac{\gamma}{2} \rightarrow 
   \frac{1}{2} A^{SU(2),3}_\phi \sigma_3 = -\frac{\gamma}{2} 
   \sigma_3 \ . 
\ee 
Thus the $SU(2)_L$ gauge field has a non-zero component $A^3_\phi$, 
\be 
   A^3_\phi = -\gamma \ . 
\ee 
Then the solutions to the Killing spinor equations (\ref{su2l}) are 
the two $SU(2)_L$ doublet $(\varepsilon_r ,\varepsilon^*_r)^T$, where 
$\varepsilon_r$ is the solution (\ref{itspinor}) and $\varepsilon^*_r$ is its 
complex conjugate. Note that the complex conjugate structure is 
consistent with the $\sigma_3$ having opposite sign diagonal entries. Note 
also that the number of Killing spinors is doubled in each sector, 
because of the label $r$. 
 
Similar manipulations are done on the $SU(2)_R$ sector. However, there 
is a subtlety when the $L$ and $R$ sector Killing spinor equations are 
combined. The two sectors each have their own $SU(2)$ gauge fields 
$A,A'$ and Killing spinor equations (\ref{su2l}), (\ref{su2r}). For 
the charged point mass spacetimes, the two background gauge fields 
need not be equal.  In general,
\be 
   A^3_\phi = -\gamma \neq  A'^3_\phi = -\gamma' \ . 
\ee 
For the point masses, the maximum supersymmetry is obtained by  setting $A = \pm A'$.
The  point mass and zero mass black hole spacetimes then have four Killing 
spinors in each sector, and the pure $\ads{3}$ background without a 
Wilson line has the maximum, eight, in each sector. Thus, as in the 
$(2,0)$ supergravity, the point masses break half of the 
supersymmetry. 
 
In summary, the supersymmetric solutions are given by a 
three-dimensional metric  
\begin{equation} 
ds^2 = - \left( {r^2 \over \ell^2} + \gamma^2 \right) dt^2 + \left( 
{r^2 \over \ell^2} + \gamma^2 \right)^{-1} dr^2 + r^2 d\phi^2 
\end{equation} 
and gauge fields 
\begin{equation} 
A_\phi^3 = \pm A_\phi^{3'} = - \gamma. 
\end{equation} 
This gives a six-dimensional metric by the Kaluza-Klein ansatz 
(\ref{metric}), which satisfies the six-dimensional equations of 
motion and preserves half the supersymmetry. In the next section, we 
will discuss how this metric arises in the near-horizon limit of the 
rotating black string.

 
\section{Conical defects from the spinning black string} 
\label{sec:spin} 
 
In the previous section, we saw how the three-dimensional solutions in
which we are interested  arose by spontaneous compactification of the  
six-dimensional ${\cal N}=4b$ theory.   
Interest in 
the six-dimensional theory is often focussed on its black string 
solutions, so we would like to see if we can relate the point 
particles to these black strings. The presence of non-trivial 
Kaluza-Klein gauge fields in the supersymmetric point particle 
solutions suggests we should consider a rotating black string, as the gauge 
field arises from off-diagonal components of the higher-dimensional 
metric and $B$-field, which we would associate with rotation. 
 
The solution describing a non-extremal spinning black string in six 
dimensions 
is~\cite{cvetic:rotint,cvetic:nearhor}\footnote{Notice that
in \cite{cvetic:rotint}, there is also a nontrivial three-form field
in the solution. We expect that this three-form
reduces, in the near-horizon limit, to our KK ansatz for 
the three-form, but we have not checked this explicitly.}
\begin{eqnarray} 
ds_6^2 = {1 \over \sqrt{H_1 H_2}} &\ & 
\left[ -(1-{2mf_D\over r^2}) d\tilde{t}^2 + d\tilde{y}^2 + H_1 H_2 
f_D^{-1} {r^4 \over (r^2 + l_1^2) (r^2 + l_2^2) - 2mr^2} dr^2 \right. 
\nonumber \\ 
&\ & 
- {4 m f_D \over r^2} \cosh\delta_1 \,\cosh\delta_2  \, (l_2 \, 
\cos^2\theta  d\psi + l_1 \, \sin^2\theta \, d\phi) d\tilde{t} 
\nonumber \\ 
&\ & 
- {4 m f_D \over r^2} \sinh\delta_1 \,\sinh\delta_2  \, (l_1 \, 
\cos^2\theta  d\psi + l_2 \, \sin^2\theta \, d\phi) d\tilde{y} 
\nonumber \\ 
&\ & 
\left( (r^2 + l_2^2 ) H_1 H_2 + 
(l_1^2 - l_2^2) \cos^2\theta ({2m f_D \over r^2})^2 \sinh^2\delta_1 \, 
\sinh^2\delta_2 \right) \cos^2\theta \, d\psi^2 
\nonumber \\ 
&\ & 
\left( (r^2 + l_1^2) H_1 H_2 + 
(l_2^2 - l_1^2) \sin^2\theta ({2m f_D \over r^2})^2 \sinh^2\delta_1 \, 
\sinh^2\delta_2 \right) \sin^2\theta \, d\phi^2 
\nonumber \\ 
&\ & 
\left. {2m f_D \over r^2} (l_2 \,  \cos^2\theta \, d\psi  + l_1 \, 
\sin^2\theta \, d\phi)^2 + H_1 H_2 r^2 f_D^{-1} d\theta^2 \right], 
\label{themetric} 
\end{eqnarray} 
where 
\begin{equation} 
H_i = 1 + {2 m \, f_D \, \sinh^2\delta_i \over r^2} 
\end{equation} 
for $i=1,2$, 
\begin{equation} 
{r^2 \over f_D} = r^2 + l_1^2 \cos^2\theta + l_2^2 \sin^2\theta, 
\end{equation} 
and $\tilde{t}$ and $\tilde{y}$ are boosted coordinates, 
\begin{equation} 
\tilde{t} = t \, \cosh \delta_0 - y \, \sinh \delta_0 \, , \, 
\tilde{y} = y \, \cosh \delta_0 - t \, \sinh \delta_0. 
\end{equation} 
For this metric, the asymptotic charges are 
\begin{eqnarray} 
M &=& m \sum_{i=0}^2 \cosh 2\delta_i, \label{Mdef} \\ 
Q_i &=& m \sinh 2\delta_i ~~~~;~~~ i=0,1,2, \\ 
J_{L,R} &=& m (l_1 \mp l_2) (\prod_{i=0}^2 \cosh\delta_i \pm 
\prod_{i=0}^2 \sinh\delta_i). 
\end{eqnarray} 

\subsection{Near-horizon limit} 
 
Cveti\v{c} and Larsen~\cite{cvetic:nearhor} showed that this metric 
has a near-horizon limit of the form BTZ $\times S^3$. To reach this 
limit, we take $\alpha' \to 0$ while holding 
\begin{equation} \label{nhlimit} 
{r \over \alpha'}, \quad {m \over \alpha'^2}, \quad {l_{1,2} \over 
\alpha'}, \quad {Q_{1,2} \over \alpha'}, \mbox{ and } \delta_0 
\end{equation} 
fixed. The resulting metric (after removing an overall factor of 
$\alpha'$) can be written as 
\begin{eqnarray} 
ds_6^2 &=& -N^2 \, d\tau^2 + N^{-2} \, d\rho^2 
+ \rho^2 \, (d\varphi - N^{\phi} \, d\tau)^2 + 
\ell^2 \, d\tilde{\Omega}_3^2, \label{btzmet}  \\
d\tilde{\Omega}_3^2 &=&
d\theta^2 + \cos^2\theta \, d\tilde{\psi}^2 + \sin^2\theta \, d\tilde{\phi}^2
\end{eqnarray} 
where 
\begin{equation} 
N^2 = {\rho^2 \over \ell^2} - M_3 +  {16 G_3^2 J_3^2 \over 
\rho^2,} 
\label{Ndef} 
\end{equation} 
\begin{equation} 
N^\phi = {4G_3 J_3 \over \rho^2}, 
\end{equation} 
and there is a non-trivial transformation between the coordinates 
$(\theta,\tilde{\phi},\tilde{\psi})$ on the near-horizon $S^3$ and the 
asymptotic coordinates, 
\begin{eqnarray} 
d\tilde{\phi} &=& 
d\phi - {R_y \over \ell^2} (l_2 \cosh \delta_0 - l_1 
\sinh \delta_0)  d\varphi - {R_y \over \ell^3} ( l_1 \cosh 
\delta_0 - l_2 \sinh \delta_0) d\tau \nonumber 
\\ 
d\tilde{\psi} &=& d\psi - {R_y \over \ell^2} (l_1 \cosh \delta_0 - l_2 
\sinh \delta_0)  d\varphi - {R_y \over \ell^3} ( l_2 \cosh 
\delta_0 - l_1 \sinh \delta_0) d\tau \, . 
\label{tilde} 
\end{eqnarray} 
The parameters of this near-horizon metric are related to the 
parameters of the full metric by 
\begin{equation} 
M_3 = {R_y^2 \over \ell^4} [(2m - l_1^2 - l_2^2) \cosh 2\delta_0 + 2 
\, l_1 \, l_2 \, \sinh 2\delta_0 ], 
\end{equation} 
\begin{equation} 
8G_3 J_3 =   {R_y^2 \over \ell^3} [ (2m - l_1^2 -l_2^2) \sinh 2\delta_0 
+ 2 \, l_1 \, l_2 \, \cosh 2\delta_0], \label{j3def} 
\end{equation} 
and $\ell = (Q_1 Q_2)^{1/4}$. The BTZ coordinates are given by 
\begin{equation} 
\tau = {t\ell \over R_y}, 
~~~~~ \varphi = {y \over R_y}, 
\label{tbtzdef} 
\end{equation} 
and 
\begin{equation} 
\rho^2 = {R_y^2 \over \ell^2} 
[r^2 + (2m - l_1^2 - l_2^2) \sinh^2\delta_0 + 2 \, l_1 \, l_2 \sinh\delta_0 
\, \cosh\delta_0]. 
\label{rbtzdef} 
\end{equation} 

The near-horizon metric looks like the direct product of a rotating 
BTZ metric and an $S^3$. However, in the original spacetime, we 
identified $\varphi \sim \varphi + 2\pi$ at fixed $\psi, \phi$, which 
is not in general the same as $\varphi \sim \varphi + 2\pi$ at fixed 
$\tilde{\psi}, \tilde{\phi}$. Thus, the coordinate transformation 
(\ref{tilde}) is not globally well-defined; that is, there are still 
off-diagonal terms in the near-horizon metric, which give rise to 
gauge fields in the three-dimensional solution. (The part of the 
transformation (\ref{tilde}) involving $\tau$ is well-defined, as 
$\tau$ is not identified.) 
 
It is convenient to trade the $l_{1,2}$ for parameters 
$a_{1,2}$ which are related to the strength of the Kaluza-Klein gauge field: 
\begin{equation} 
a_1 = l_1 \cosh \delta_0 - l_2 \sinh \delta_0, \quad a_2 = l_2 \cosh 
\delta_0 - l_1 \sinh \delta_0. 
\end{equation} 
Then we can write 
\begin{equation} 
\tilde{\phi} = \phi - {R_y \over \ell^2} \, a_2 \varphi - 
{R_y \over \ell^3} \, a_1 \tau, \quad \tilde{\psi} = \psi - {R_y 
\over \ell^2} \, a_1 \varphi - {R_y \over \ell^3} \, a_2 
\tau, 
\end{equation} 
and the relations between the near-horizon and full metric parameters 
become 
\begin{equation} 
8G J_3 = {R_y^2 \over \ell^3} (2m \sinh 2\delta_0 + 2 a_1 a_2) 
\end{equation} 
and 
\begin{equation} 
M_3 = {R_y^2 \over \ell^4} (2m\cosh 2\delta_0 - a_1^2 - a_2^2). 
\label{m3def} 
\end{equation} 
It is more convenient to keep some $l_2$ dependence in $\rho$, 
and write it as 
\begin{equation} 
\rho^2 = {R_y^2 \over \ell} (r^2 + 2m \sinh^2 \delta_0 
+l_2^2 - a_2^2). 
\end{equation} 
To extract the Kaluza-Klein gauge fields, we need to write the metric
on the 3-sphere in the coordinates used in Sec.~\ref{sec:kk}.   This
coordinate transformation is given in App.~\ref{sphereapp}.  The
result is:
\begin{equation} 
A^3 = {R_y \over \ell^2} (a_1-a_2) d\varphi, \quad A^{3'} = - {R_y \over 
\ell^2} (a_1+a_2) d\varphi, 
\end{equation} 
where the indices $3$, $3'$ refer to $SU(2)_L$ and $SU(2)_R$ 
respectively. The near-horizon limit of the spinning black string thus 
gives a three-dimensional metric of BTZ form coupled to gauge 
fields. Furthermore, the BTZ mass $M_3$ (\ref{m3def}) can be negative 
for suitable choices of the parameters (in particular, it is possible 
to make $M_3$ negative while $m \geq 0$).

We can now choose the parameters so that we recover the supersymmetric 
point particle solutions of the preceding section. For simplicity, we 
have only considered non-rotating conical defects, so we require 
$J_3=0$. Since we seek a supersymmetric 
solution, it's reasonable to set $m=0$. Then $J_3=0$ implies $a_1 
a_2=0$; without loss of generality, take $a_2=0$. Note that for this 
choice of parameters, all dependence on $\delta_0$ disappears from the 
metric. The mass and gauge field are now 
\begin{equation} 
M_3 = -{R_y^2 \over \ell^4} a_1^2  \equiv -\gamma^2
\end{equation} 
and 
\begin{equation} 
A^3 = -A^{3'} = {R_y \over \ell^2} a_1 d\varphi = \gamma \, d\phi \, . 
\end{equation} 
Therefore, we recover the conical defects of the previous section.
 
The near-horizon limit of strings with physically reasonable choices
for the parameters can thus give rise to point particle spacetimes,
with negative values for $M_3$. Remarkably, this shows that {\it
global} $\ads{3}$ appears as the near-horizon limit of a suitable
compactified black string.\footnote{The Wilson line that appears in this
limit of our solutions can be removed by a coordinate transformation
from the 6d point of view.} To explore the consequences of this, it
will be useful to also consider a family of non-extremal solutions
with the same parameters. A convenient choice is to take $\delta_0=0$,
$a_2=0$ (which is equivalent to $\delta_0 = 0$, $l_2=0$). In this
case, $J_3=0$ and $M_3 = R_y^2 (2m-a_1^2)/\ell^4$.
 
\subsection{The full metric} 
 
Having seen that point particles can arise in the near-horizon limit
of spinning black strings, we would like to be able to say something
about the geometry of the full string solution.  The near-horizon
limit is also a near-extreme limit of the full black string.  The
extremal limit involved is\footnote{Note that this implies $Q_0 \to
0$, and is hence not the same as the limit $m \to 0$ with $Q_{0,1,2}$
fixed that is usually considered in the context of studies of extremal
black strings
\cite{breckenridge:spin}.} 
\begin{equation} 
m \to 0, \quad Q_{1,2} \mbox{ and } \delta_0 \mbox { fixed.} \label{limit} 
\end{equation} 
Initially, we will leave the value of 
$a_2$ unspecified. In this limit, 
\begin{eqnarray} 
M &=& Q_1 + Q_2, \\ J_{L,R} &=& {\sqrt{Q_1 Q_2} \over 2} (l_1 \mp 
l_2) (\cosh \delta_0 \pm 
\sinh\delta_0) = {\sqrt{Q_1 Q_2} \over 2}(a_1 \mp a_2). 
\end{eqnarray} 
The coordinate transformation $\bar{\rho}^2 = r^2 + l_2^2 - 
a_2^2$ results in an  extremal metric in the extremal metric of the
form 
\begin{eqnarray} 
ds_6^2 = {1 \over \sqrt{H_1 H_2}} &\ & 
\left[ -dt^2 + dy^2 + H_1 H_2 
g_D^{-1} {\bar{\rho}^4 \over (\bar{\rho}^2 + a_1^2) (\bar{\rho}^2 + 
a_2^2)} d\bar{\rho}^2 \right. \nonumber \\ 
&\ & 
- {2 \sqrt{Q_1Q_2} g_D \over \bar{\rho}^2} [\cos^2 \theta d\psi (a_2 dt 
+ a_1 dy) + \sin^2 \theta d\phi (a_1 dt + a_2 dy)] 
\nonumber \\ 
&\ & 
+\left( (\bar{\rho}^2 + a_2^2) H_1 H_2 + 
(a_1^2 - a_2^2) \cos^2\theta ({g_D \over \bar{\rho}^2})^2 Q_1 
Q_2 \right) \cos^2\theta \, d\psi^2 
\nonumber \\ 
&\ & 
+\left( (\bar{\rho}^2 + a_1^2) H_1 H_2 + 
(a_2^2 - a_1^2) \sin^2\theta ({g_D \over \bar{\rho}^2})^2 Q_1 
Q_2 \right) \sin^2\theta \, d\phi^2 
\nonumber \\ 
&\ & 
+\left. H_1 H_2 \bar{\rho}^2 g_D^{-1} d\theta^2 \right], 
\label{metricc2} 
\end{eqnarray} 
where 
\begin{equation} 
H_i = 1 + {g_D Q_i \over \bar{\rho}^2} 
\end{equation} 
for $i=1,2$, and 
\begin{equation} 
{\bar{\rho}^2 \over g_D} = \bar{\rho}^2  + a_1^2 \cos^2\theta + a_2^2 
\sin^2\theta. 
\end{equation} 
The metric is now independent of $\delta_0$. That is, when we take the 
extremal limit with $\delta_0$ fixed, we find that it becomes just a 
coordinate freedom in the limit.  This is presumably a form of the 
usual restoration of boost-invariance at extremality. Thus, the fact 
that the near-horizon extremal metric did not depend on this parameter 
is a property of the extremal limit, not the near-horizon limit. If we 
take $a_2=0$, we find that $J_L = J_R$. 
 
We can also consider the non-extremal metric with $\delta_0=0$, 
$l_2=0$ (corresponding to the simple family of non-extremal 
generalizations we considered in the previous section). The form of 
the metric is not substantially simplified relative to (\ref{metric}), 
so we will not write it out again here. We merely note that this 
metric has  a single horizon at $r^2 = 2m-l_1^2$, of area 
\begin{equation} 
A = 8\pi^3 m R_y \cosh \delta_1 \cosh \delta_2 \sqrt{2m-l_1^2}. 
\end{equation} 
In the near-horizon limit, this reduces to $2 \pi \ell \sqrt{M_3} 
\times 4\pi^2 \ell^3$, which we recognize as the product of the 
area of the BTZ black hole horizon and the volume of the $S^3$, as 
expected.

\subsection{Properties of the solution: Instabilities and singularities} 

From the three-dimensional point of view, there is a conical
singularity at $\rho=0$, for both the non-rotating BTZ black holes and
for the point particle spacetimes. In the full six-dimensional
solution, we need to check the nature of this singularity. The
curvature invariants are everywhere finite, so there is no curvature
singularity. Consider a small neighborhood of the point $\rho=0, \, 
\theta=0$ in a constant 
time slice. The metric near this point can be approximated by
\begin{equation}
ds^2 \approx {d\rho^2 \over \gamma^2} + \rho^2 d\varphi^2 + d\theta^2 +
d\phi^2 + \theta^2 (d\psi + \gamma d \varphi)^2 \, .
\end{equation}
This suggests a further coordinate transformation 
\begin{equation}
\rho = \gamma R \cos u \ , \theta  = R \sin u \ ,
\end{equation}
which brings the metric to the form
\begin{equation}
ds^2 \approx dR^2 + R^2 (du^2 + \gamma^2 \cos^2 u d\varphi^2 + \sin^2
u (d\psi + \gamma d\varphi)^2).
\end{equation}
Thus, the area of a surface at $\epsilon$ proper distance from the
point $\rho=0, \theta=0$ is $\epsilon^3 \gamma 2\pi^2$. The difference
between this area and the standard $\sphere$ area $\epsilon^3 2\pi^2$
indicates that there is a conical defect at this point. Note that the
choices of parameters for which we get negative $M_3$, and hence a
point particle solution, are precisely those for which the full
six-dimensional solution does not have an event horizon.  Hence this
is a naked conical singularity.
 
For a given value of $a_1$, we can obtain point particle solutions 
with all values of $M_3$ by varying $R_y$. There is no obvious bound 
associated with the value $M_3 = -1$ corresponding to pure AdS 
space. It was already noted by Izquierdo and Townsend in 
\cite{Izq:2+1} that there exist supersymmetric solutions to 3d gravity
for arbitrarily negative values of $M_3$. These solutions are all
singular, and the singularities which occur for $M_3 < -1$ are not
essentially different from those which occur for $M_3 > -1$. From a
three-dimensional point of view, one simply asserts that while the
singular solutions with $M_3 >-1$ are physically relevant, as they can
arise from the collapse of matter, those with $M_3 < -1$ are
physically irrelevant. We similarly expect that only the solutions
with $M_3 > -1$ will have a physical interpretation in the dual CFT,
as AdS space corresponds to the NS vacuum of the CFT, and we do not
expect to find excitations with lower energy. It is therefore
surprising is that the six-dimensional string metric makes no
distinction between $M_3 < -1$ and $M_3 > -1$. It is clear that it
does not, as the nature of the singularity in the six-dimensional
solution is independent of the value of $R_y$.

However, we should still ask whether this solution is stable for all
values of $R_y$.  In~\cite{peet:micro}, it was argued the ${\rm BTZ}
\times S^3$ solution (for all masses) would be stable against
localization on $\sphere$ so long as global $\ads{3}$ did not appear
in the spectrum of the compactified string.  Here we have argued that
for certain parameters, the rotating, compactified string does include
global $\ads{3}$.  Therefore, it is doubly worthwhile to consider the
question of instabilities for near-extremal solutions with angular
momentum.\footnote{It was argued in~\cite{peet:micro} that such a
localization instability should not occur for the full asymptotically
flat black string solutions, as it would break spherical symmetry. In
our case, the spherical symmetry is already broken by the rotation; so
it is not obvious that this argument applies.}

In fact, the full asymptotically flat rotating black string solution
has a more familiar instability: localization on the circle ($y$) 
along which the string is compactified.  Such an instability 
typically sets in 
when the entropy of the localized solution is greater than that of the
extended one~\cite{greglaff}.  Since the present solution carries a charge, a simple
model for the localized solution is the extreme black string carrying
the same charge, along with a six-dimensional Schwarzschild black hole
carrying the energy above extremality of the original
solution. Consider, for definiteness, the non-extremal solutions
discussed above, with $\delta_0 = 0$, $l_2 = 0$. From (\ref{Mdef}), $M
- M_{ext} \approx m R_y$ for near-extremal solutions, so the entropy
of the Schwarzschild black hole in the candidate localized solution is
\begin{equation} 
S_{BH} \sim (m R_y)^{4/3}. \label{sgain} 
\end{equation} 
Thus, for $R_y > R_{crit}$, we expect the solution to be unstable, 
where $R_{crit}$ is given by $S_{BS} = S_{BH}$.   That is, 
\begin{equation} \label{Rcrit} 
R_{crit}^{2/3} \sim {Q_1 Q_2 (2m - l_1^2) \over m^{8/3}} 
\end{equation} 
for near-extremal solutions. Thus, as we approach extremality, 
$R_{crit}$ may grow, but it will eventually decline 
and reach zero at $m = l_1^2/2$. For fixed $R_y$, all the 
near-extremal solutions with $m$ small enough are unstable to 
localization.\footnote{This is quite different from the usual 
behavior near extremality: for a non-rotating black string, $R_{crit} 
\to \infty$ as $m \to 0$, as we can see from (\ref{Rcrit}) with 
$l_1=0$.} This instability sets in at a finite distance from
extremality;  so we will always encounter it before reaching the 
instability to localization on $\sphere$ that is suggested by the physics of
the near-horizon limit.
 
There is hence an $R_y$-dependent instability. Does this allow us to
exclude the undesirable singularities (those with $M_3 <-1$)?  We have
argued for this instability by comparing the entropy of a near-extreme
string to that of the extreme string plus a localized black hole. Thus
we have {\it assumed} that the extreme string, which corresponds to a
supersymmetric point particle solutions, is stable, and we cannot use
this approach to argue that the extremal solutions are unstable.  The
assumption of stability of the extremal solutions is consistent,
since, as we approach extremality, the entropy gain in the
localization (\ref{sgain}) is going to zero.  Furthermore, there
is no lower-energy system than the extreme string that carries the
same angular momentum and charges. Together with experience in other
examples, this suggests the extreme string is stable for all values of
$R_y$, and hence instabilities do not serve to rule out the cases
corresponding to $M_3 < -1$.

\section{A proposal for a dual description} 
\label{sec:dual} 
 
In~\cite{david:point}, an interpretation of the point mass geometries
in terms of spectral flow operators was given.  Here, we 
propose a somewhat different model in terms of density
matrices in the RR sector of the boundary CFT. It may seem surprising
to propose that a gravitational system without a horizon, and hence no
Bekenstein-Hawking entropy, would be described by a density matrix.
However, the classical formulae only register a sufficiently large
degeneracy.  The ensemble of supersymmetric states that we are proposing
contains fewer states than the number that enter the ensemble
describing the $M=0$ black hole.  As is well known, the latter system
has vanishing entropy in the semiclassical limit. Below, we briefly
summarize the main idea of our proposal. Details and various tests 
will be presented in a future publication~\cite{uslater}.
 
All geometries we have considered are either singular or have an 
horizon. Once we remove the singular region, we are left with 
a space with topology $R^2\times S^1$. This is true even for 
pure  $AdS_3$ with  nonzero $SU(2)$ Wilson lines. The singularity 
in those cases is not a curvature singularity, but one where 
the $SU(2)$ gauge fields are ill-defined. The only exception 
is pure $AdS_3$ without Wilson lines, whose topology is that 
of $R^3$. We will first ignore pure $AdS_3$, but as we will 
see a bit later it fits in quite naturally. 
 
On a space with topology $R^2 \times S^1$, there are two 
topological choices for the spin bundle, corresponding 
to periodic and anti-periodic boundary conditions along 
the $S^1$. By periodic and anti-periodic we refer to 
spinors expressed in terms of a Cartesian frame on the 
boundary cylinder, which correspond to a radial frame 
in the $AdS$ geometry. Thus, periodic boundary conditions 
correspond to the RR sector, anti-periodic boundary 
conditions to the NS sector. The proposed dual description of 
the point mass geometries will be valid assuming periodic 
boundary conditions, but as we will see, one can 
derive an equivalent 
description using anti-periodic boundary conditions. 
 
It may be confusing that we impose periodic boundary 
conditions on the spinor and fermion fields, because 
if we use the field equations to parallel transport 
a spinor  along the circle, we can pick up arbitrary 
phases, depending on the choice of point mass geometry, 
and also on the choice of $SU(2)$ Wilson lines. 
These phases are the holonomies of the flat  $SL(2)$ 
and $SU(2)$ connections that define  the geometry 
and Wilson  lines, but they are still connections on 
the same topological spinor bundle. In other words, 
given a bundle with a given topology, there are still 
many flat connections on that bundle, which 
are parameterized by its holonomies. In our case we 
choose the (periodic) spinor bundle, and view the 
gauge fields as connections on this bundle. Whether 
there exist global covariantly constant sections of 
the spinor bundle is a question that does depend 
crucially on the choices of flat connections, and 
is precisely the question whose answer tells us whether or not
a  given solution preserves some supersymmetries.

The near horizon geometries in Sec.~\ref{sec:spin}, that include 
the BTZ and spinning point particle solutions, 
depend on five quantities, namely 
$\ell=(Q_1 Q_2)^{1/4}$, $M_3$, $J_3$, $A^3$, $A^{3\prime}$. 
In order to give the dual conformal field theory 
description, we define 
\begin{eqnarray} 
c & = & \frac{3\ell}{2G_3} \\ 
\ell_0 & = & \frac{\ell M_3 + 8 G_3 J_3}{16 G_3} \\ 
\bar{\ell}_0 & = & \frac{\ell M_3 - G_3 J_3}{16 G_3} \\ 
j_0 & = & \frac{c}{12} A^3 \\ 
\bar{j}_0 & = & \frac{c}{12} A^{3'}  . 
\end{eqnarray} 
Our proposal is that the geometry corresponds in the 
boundary theory to a density matrix of (equally weighted) 
states in the RR sector with quantum numbers 
\begin{eqnarray} 
J_0 & = & j_0 \label{xxx1} \\ 
\bar{J}_0 & = & \bar{j}_0  \\ 
L_0 & = & \ell_0 + \frac{c}{24} + \frac{6(j_0)^2}{c} 
\label{xxx3} \\ 
\bar{L}_0 & = & \bar{\ell}_0 + 
\frac{c}{24} + \frac{6(\bar{j}_0)^2}{c} . \label{xxx4} 
\end{eqnarray} 
 
The quadratic terms in $L_0$ and $\bar{L}_0$ may appear 
surprising, but there are several ways to justify them. 
First of all, in this way $\ell_0$ and $\bar{\ell}_0$ 
are spectral flow invariants, and the asymptotic density 
of RR states with the quantum numbers (\ref{xxx1})--(\ref{xxx4}) 
is a function of $\ell_0,\bar{\ell}_0$ only. This is in 
nice agreement with the fact that the area of the horizon 
and therefore the entropy of BTZ black holes also 
depends on $\ell_0,\bar{\ell}_0$ only. 
 
The quadratic terms in (\ref{xxx3}) and (\ref{xxx4}) are also 
natural if we use the relation between the Hamiltonian 
reduction of $SU(1,1|2)$ current algebra and the boundary 
superconformal algebra \cite{hamred1,hamred2,hamred3,janads,hamred4}. 
The stress tensor obtained in this 
Hamiltonian reduction procedure contains the Sugawara stress tensor 
of the $SU(2)\subset SU(1,1|2)$ current algebra, and this 
extra contribution yields the quadratic terms in (\ref{xxx3}), 
(\ref{xxx4}). 
 
Spectral flow in the boundary theory corresponds in the 
bulk to the following procedure. In the bulk, we can 
remove part of the $SU(2)$ Wilson lines by a singular field 
redefinition. Namely, if a field $\psi(x)$ has charge $q$ 
under the $U(1) \subset SU(2)$ subgroup, we can introduce 
new fields 
\begin{equation} 
\tilde{\psi}(x)  = P \exp ( q \xi \int_{x_0}^x A \cdot dx) 
\psi(x) 
\end{equation} 
and at the same time replace the gauge field by 
\begin{equation} 
\tilde{A}(x) = (1-\xi) A(x) . 
\end{equation} 
This is a (singular) gauge transformation and does not 
affect the physics. The only consequence of this transformation 
is that it gives twisted boundary conditions to all fields 
charged under the $U(1)$. If we compute the new quantum 
numbers according to (\ref{xxx1})-(\ref{xxx4}), we find 
\begin{eqnarray} 
J'_0 & = & J_0(1-\xi)  \\ 
L'_0 & = & L_0 -\frac{12}{c} \xi J_0^2 + \frac{6}{c} \xi^2 J_0^2 
\end{eqnarray} 
which is precisely the behavior of these quantum numbers under 
spectral flow with parameter $\eta=\frac{12}{c} \xi j_0$~\cite{specflow}. 
In other words, we can set up the AdS-CFT correspondence with 
arbitrary twisted boundary conditions. The twisted boundary 
conditions in the bulk match the twisted boundary conditions 
of the CFT, and the relations (\ref{xxx1})--(\ref{xxx4}) are 
valid independently of the twist. Spectral flow corresponds 
to a field redefinition both in the bulk and in the boundary 
theory, and does not affect the physics.  For other discussions of the
role of spectral flow, see~\cite{specflow1,specflow2,david:point,hamred4}.
 
We can now understand how pure AdS arises in this picture. 
We start with pure AdS with a flat gauge field with holonomy $-1$ 
in the fundamental representation. According to the above 
proposal, this corresponds to states in the RR sector with 
$L_0=c/24$ and $J_0=c/12$. If we remove the gauge field 
completely by a field redefinition, this changes the boundary 
conditions of the fermions, and they become anti-periodic 
instead of periodic. Therefore, the field redefinition brings 
us from the R to the NS sector. In addition, the quantum numbers 
after the field redefinition become $L_0=J_0=0$. We see 
that pure AdS with anti-periodic boundary conditions (the only 
boundary conditions that are well-defined on pure AdS) 
corresponds to the vacuum in the NS sector, as expected. 
 
As a final check of our proposal, we will rederive the results 
of Izquierdo and Townsend \cite{Izq:2+1} regarding the supersymmetries 
in point mass geometries with non-trivial gauge fields turned on. 
Consider again the point mass geometries with $M_3=-\gamma^2$, 
and $J_3=0$, and only look at the left moving sector. The 
equation for $L_0$ reads 
\begin{equation} 
L_0 = (1-\gamma^2)\frac{c}{24}+ \frac{c}{24} (A^3)^2 
\end{equation} 
where $A$ is the value of the $U(1)_L$ gauge field. 
The two choices of spin bundle give two inequivalent 
situations. If we take periodic boundary conditions for the fermions, 
we find a state with 
\begin{equation} 
J_0 = \frac{c}{12} A^3, \qquad 
L_0 = (1-\gamma^2)\frac{c}{24}+ \frac{c}{24} 
(A^3)^2 
\label{qn1} 
\end{equation} 
in the RR sector. If  we start with anti-periodic boundary 
conditions for the fermions we find a state with 
quantum numbers (\ref{qn1}), but now in the NS sector. 
Using the spectral flow procedure outlined above, this can 
be mapped to a state in the RR sector with 
\begin{equation} 
J_0 = \frac{c}{12}(A^3+1), \qquad 
L_0 = (1-\gamma^2)\frac{c}{24}+ \frac{c}{24} 
(A^3+1)^2
\label{qn2} 
\end{equation} 
There are also spectral flows that map the RR sector to itself, 
and these are labeled by an integer $n$. Applying these spectral 
flows to (\ref{qn1}) we obtain states in the RR sector with 
\begin{equation} 
J_0 = \frac{c}{12}(A^3+2n) , \qquad 
L_0 = (1-\gamma^2)\frac{c}{24}+ \frac{c}{24} 
(A^3 + 2 n)^2
\label{qn3} 
\end{equation} 
and from (\ref{qn2}) we obtain states with 
\begin{equation} 
J_0 = \frac{c}{12} (A^3 + 2 n + 1), \qquad 
L_0 = (1-\gamma^2)\frac{c}{24}+ \frac{c}{24} 
(A^3 + 2 n + 1)^2 .
\label{qn4} 
\end{equation} 
The quantum numbers in (\ref{qn3}) and (\ref{qn4}) can be summarized 
by the equations 
\begin{equation} 
J_0 = \frac{c}{12} (A^3 + n), \qquad 
L_0 = (1-\gamma^2)\frac{c}{24}+ \frac{c}{24} 
(A^3 + n)^2
\label{qn5} 
\end{equation} 
where $n$ is an arbitrary integer. In the RR sector, supersymmetry is 
preserved for RR ground states with $L_0=c/24$ only. Thus, we need 
that 
\begin{equation} 
A^3 = \pm \gamma +n 
\end{equation} 
for some integer $n$. This is precisely the same condition as found 
in~\cite{Izq:2+1}, see equations (\ref{jj8}) and (\ref{jj9}).

 
\section{Summary and discussion} 
\label{sec:disc}

We have embedded the 3d BPS conical defects into a higher dimensional
supergravity arising from string theory.    The defects in three
dimensions provide particularly simple laboratories for the AdS/CFT
correspondence.    They are examples of systems that are neither
perturbations of the AdS vacuum, nor semiclassical thermal states like
black holes.   Understanding the detailed representation of such
objects in a dual CFT is bound to be instructive.    Furthermore, the
conical defects which we have constructed in six dimensions can be
collided to yield the (near horizon limit) of the classic 5d black
holes whose entropy was explained by Strominger and
Vafa~\cite{stromvaf}.

To recap, we have given a detailed analysis of the Kaluza-Klein
reduction of the ${\cal N} = 4b$ chiral supergravity in six dimensions
coupled to tensor multiplets.  Our KK ansatz gives solutions to the 6d
equations of motion which correspond from the dimensionally reduced point of
view to 3d conical defects with Wilson lines.  Supersymmetry is
preserved by a judicious choice of the gauge potential.  From the 6d
point of view, our solutions are spheres fibered over an $\ads{3}$
base, and the conical defect arises at a point where the fibration
breaks down.   Although we thereby embed all the solutions of the 3d
Chern-Simons supergravities into the six dimensional theory, our
ansatz does not in general produce a consistent truncation to a
Chern-Simons theory.  (Solutions with $F=0$ are admitted, but the six
dimensional equations of motion do not impose this.)\footnote{While
this paper was in the final stages of preparation we became 
aware that related investigations have been conducted by Samir Mathur.}

Our solutions can also be understood as near-horizon limits of
rotating string solutions in six dimensions compactified on a circle.
Surprisingly, global $\ads{3} \times S^3$ appears in one corner of the
parameter space.  Although our solutions contain conical
singularities, they remain interesting because we expect them to be
resolved by string theory.  In particular, we have a proposal for a
non-singular dual description in a conformal field theory.  If our
solutions are admissible, they appear to imply an Gregory-Laflamme
instability for the near-extremal rotating black strings.

We have suggested a concrete representation of our conical defects as
ensembles of chiral primaries in a dual CFT.  Subsequent articles will
test our proposal.

\vskip0.1in {\leftline {\bf Acknowledgments}} 
 
We are particularly grateful to Ray Tomlinson of ARPANET, who invented
e-mail~\cite{email}, the Wright brothers, and Alexander Graham Bell,
who made this project possible.  In addition, we have benefited from
conversations with Sander Bais, Mirjam Cveti\v{c}, Sumit Das, Robbert
Dijkgraaf, Per Kraus, Don Marolf, Samir Mathur, Andy Strominger, and
Arkady Tseytlin. During the course of this work {\small VB} was
supported first by the Harvard Society of Fellows and the Milton Fund
of Harvard University, and then by DOE grant DOE-FG02-95ER40893.  The
authors thank the Aspen Center for Physics, Harvard University and
University of Amsterdam for hospitality at various points in this
project.

\appendix 
 
\section{From 6d symplectic Majorana spinors to 3d spinors} 
\label{sympapp} 
In this appendix, we discuss the symplectic Majorana condition on  
6d chiral spinors. In particular, we show in detail how the 6d spinors 
can be chosen to be $\su$ doublets of complex conjugate 
two-component spinors 
\be 
    \varepsilon_r = \left( \begin{array}{c} \varepsilon^{(2)_r} \\ 
     \varepsilon^{(2)*}_r \end{array} \right)  \ . 
\label{toshow} 
\ee 
 
The 6d Killing spinor equation in $N=4b$ supergravity was 
\begin{equation} \label{ksb} 
(D_M \mp \frac{1}{4} H^5_{MNP} \Gamma^{NP} ) 
\epsilon_r 
=0\ , 
\end{equation} 
where the upper (lower) sign is for $r=1,3$ ($r=2,4$). 
The supersymmetry parameters $\epsilon_r$ are positive chirality 
spinors 
\be 
         \epsilon_r = \left( \begin{array}{c} \varepsilon_r \\ 0 
         \end{array} \right) \ . 
\ee 
Each of the four ($r=1,\ldots ,4)$ spinors has four complex
components.  That gives 32 real degrees of freedom, of which we must
remove half, since the $N=4b$ supergravity has only 16
supersymmetries. This can be done by imposing a reality condition on
the chiral spinors. In 6d, the appropriate reality condition is either
the SU(2) or the symplectic Majorana condition, depending on the
R-symmetry of the supersymmetry algebra \cite{KT}. It can be
consistently imposed along with the chirality projection.  Literature
on the subject includes \cite{KT,HST,romans,deger:6d,NT}.  Here we are
mostly following \cite{HST}.
 
Ref. \cite{HST} first considers $N=2$ susy in 6d. There is an SU(2) 
doublet of four-component complex spinors, satisfying the 
SU(2)-Majorana condition 
\be 
    ( \psi^i_{\alpha} )^* \equiv \bar{\psi}_{\dot{\alpha} i} 
     = \epsilon_{ij} B^{\ \beta}_{\dot{\alpha}} \psi^j_{\beta} 
\ee 
where $i,j=1,2$ label the doublet and $\alpha,\dot{\alpha}$ are 
spinor indices. The matrix $B$ must satisfy 
\be 
    BB^* = B^*B = -1 . 
\label{BB} 
\ee 
One can see this by applying the SU(2)-Majorana condition twice 
and remembering that $\epsilon_{21}=-\epsilon_{12}=-1$. 
 
For N=4 supersymmetry, we have four complex four-component 
spinors, transforming as a fundamental of the USp(4) R-symmetry 
group. The four-component spinors can be 
understood as chiral 8-component complex spinors, with 4 
components projected out by the chirality projection. Now the 
SU(2)-Majorana condition is promoted to a symplectic Majorana 
condition 
\be 
   \bar{\Psi}_{r\dot{\alpha}} = \Omega_{rs} B^{\ \beta}_{\dot{\alpha}} 
   \Psi_{s\beta} 
\label{syma1} 
\ee 
where $\Omega_{rs}$ is the symplectic metric of the USp(4) group, 
and $\dot{\alpha},\beta$ label the 8 components of the spinor. B 
is a $4\times 4$ matrix satisfying (\ref{BB}). The symplectic 
metric is 
\be 
   \Omega = \left( \begin{array}{rr} {\bf 0} & {\bf 1} \\ 
   {\bf -1} & {\bf 0} \end{array} \right) \ . 
\label{Otrad} 
\ee 
Let us take the spinors $\Psi_r$ to be the chiral 
8-component  spinors $\epsilon_r$. 
Recall that we have chosen the spinors $\epsilon_r$ with $r=1,3$ to 
have opposite $\Gamma^5$ eigenvalues from $r=2,4$. In this choice, we 
have ensured that the symplectic metric will not mix spinors with 
opposite eigenvalues. 
 
For the supersymmetry parameters, 
the symplectic Majorana condition (\ref{syma1}) becomes 
\be 
   \bar{\epsilon}^{T}_1 = B \epsilon_3 \ , 
\label{syma2} 
\ee 
and similarly for $\bar{\epsilon}_2,\epsilon_4$. The left hand 
side of (\ref{syma2}) is 
\bea 
  \bar{\epsilon}^{T}_1 &=& (\epsilon^{\dagger}_1 \Gamma_0 )^T 
   \nonumber \\ 
    \mbox{} &=& \left( \begin{array}{cc} 0 & 1\otimes \gamma^{0,T} 
  \\ 1\otimes \gamma^{0,T} & 0 \end{array} \right) 
 \left( \begin{array}{c} \varepsilon^*_1 \\ 0 \end{array} 
  \right) \nonumber \\ 
  \mbox{} &=& \left( \begin{array}{c} 0 \\ -(1\otimes \gamma^0) 
  \varepsilon^*_1  \end{array} \right) \ , 
\eea 
where in the last line we used $\gamma^{0,T}=-\gamma^0$ (recall 
that $\gamma^0 = -i\sigma_2$). 
 
To evaluate the right hand side of (\ref{syma2}), we need the matrix $B$. 
We can assume it to be real, and of the form 
\be 
     B= \left( \begin{array}{cc} \mbox{} & \hat{B} \\ \hat{B} & 
     \mbox{} \end{array} \right) \ , 
\ee 
where $\hat{B}$ is real 4x4-matrix satisfying $\hat{B}^2 = -1$. 
A convenient choice turns out to be 
\be 
  \hat{B} = \sigma_1 \otimes \gamma^0 \ . 
   \label{Bhat} 
\ee 
The right hand side of (\ref{syma2}) becomes 
\be 
  B\epsilon_3 
 =  \left( \begin{array}{c} 0 \\ \hat{B} \varepsilon_3 
 \end{array} \right) \ . 
\ee 
Thus, (\ref{syma2}) reduces to the equation 
\be 
   -(1\otimes \gamma^0) \varepsilon^*_1 = \hat{B} \varepsilon_3 = (\sigma_1 
   \otimes \gamma^0) \varepsilon_3 \ . 
\label{syma3} 
\ee 
Next, introduce the notation 
\be 
   \varepsilon_r = \left( \begin{array}{cc} \chi_r \\ \xi_r \end{array} 
   \right) \ , \ r=1,3 
\ee 
where $\chi_r,\xi_r$ are 2-component complex spinors. Then (\ref{syma3}) 
is equivalent to 
\be 
  \left( \begin{array}{r} -\gamma^0 \chi^*_1 \\ -\gamma^0 \xi^*_1 
  \end{array} \right) = \left( \begin{array}{r} \gamma^0 \xi_3 
  \\ \gamma^0 \chi_3 \end{array} \right) \ . 
\ee 
Thus the two 4-component spinors $\varepsilon_{1,3}$ are 
\be 
  \varepsilon_1 = \left( \begin{array}{r} \chi_1 \\ \xi_1 
  \end{array} \right) \ \ ; \ \ 
  \varepsilon_3 = -\left( \begin{array}{r} \xi^*_1 \\ \chi^*_1 
  \end{array} \right) \ . 
\ee 
Out of the 8 complex degrees of freedom, only 4 remain. Since the 
Killing spinor equations are linear, we can take linear 
combinations of $\varepsilon_1,\varepsilon_3$: 
\bea 
      \tilde{\varepsilon}_1 &=& \varepsilon_1 - \varepsilon_3 \nonumber \\ 
      \tilde{\varepsilon}_3 &=& i(\varepsilon_1 +\varepsilon_3) 
\eea 
Then, the $\tilde{\varepsilon}_r$ are of the complex conjugate 
doublet form (\ref{toshow}). 
The corresponding 8-component spinors are 
\be 
       \tilde{\epsilon}_r = \left( \begin{array}{c} \tilde{\varepsilon}_r \\ 0 
  \end{array} \right) \ . 
\ee 
The same can be done to the $r=2,4$ spinors which had the opposite 
$\Gamma^5$ eigenvalues. We can then drop the tildes, 
and assume that in the Killing spinor calculation the 6d spinors are such 
that the resulting 3d 
spinors will be of the form (\ref{toshow}).

\section{The 3-sphere} 
\label{sphereapp} 
 
The 3-sphere of radius $\ell$ is explicitly described as: 
\begin{eqnarray} 
\ell^2 &=& x_1^2 + x_2^2 + x_3^2 + x_4^2  \ ,\\ 
ds^2 &=& dx_1^2 + dx_2^2 + dx_3^2 + dx_4^2 \ . 
\end{eqnarray} 
One solution to the constraint is 
\begin{eqnarray} 
x_1 &=& \ell \, \cos\theta \ ,\\ 
x_2 &=& \ell \, \sin\theta \, \cos\phi \ ,\\ 
x_3 &=& \ell \, \sin\theta \, \sin\phi \, \cos\psi \ ,\\ 
x_4 &=& \ell \, \sin\theta \, \sin\phi \, \sin\psi \ , 
\end{eqnarray} 
which gives the metric 
\begin{equation} 
ds^2 = \ell^2(d\theta^2 + s_\theta^2 \, d\phi^2 + s_\theta^2 \, s_\phi^2 \, 
d\psi^2) \ . 
\end{equation} 
(We are using the notation $s_\theta \equiv \sin\theta$ and $c_\theta 
\equiv \cos\theta$.)   The generators of the $\so$ isometry group of $S^3$
are $\Lambda^i_j \sim x^i \, \partial_j - x^j \, \partial_i$.  We are
actually interested in exposing the $\su \times \su$ structure and so
it is  better to go to complex coordinates. Let $z_1 = x_1 + i x_2$ $z_2
= x_3 + i x_4$.  Then the sphere can also be written as:
\begin{equation} 
ds^2 = dz_1 \, d\bar{z}_1 + dz_2 \, d\bar{z}_2 
~~~~~;~~~~~~ 
\ell^2 = z_1 \, \bar{z}_1 + z_2 \, \bar{z}_2 \ . 
\end{equation} 
Let us parametrize solutions to these equations as: 
\begin{eqnarray} 
z_1 &=& \ell \, \cos(\theta/2) \, e^{i(\phi + \psi)/2} \ ,\\ 
z_2 &=& \ell \, \sin(\theta/2) \, e^{i(\phi - \psi)/2} \ . 
\end{eqnarray} 
(Note that exchanging $\phi \leftrightarrow \psi$ complex conjugates 
$z_2$.)  We arrive at the $\sphere$ metric 
\begin{equation} 
ds^2 = {\ell^2 \over 4} \, [d\theta^2 + d\phi^2 + d\psi^2 + 2\cos\theta 
\, d\phi \, d\psi] \ .  
\label{metricsp} 
\end{equation}

\subsection{$\su \times \su$} 
In the complex coordinates, it is clear that there are two $\su$ 
symmetries under which $\sphere$ is invariant: 
\begin{equation} 
\pmatrix{z_1 \cr z_2} \rightarrow U_L \, \pmatrix{z_1 \cr z_2} 
~~~~~;~~~~~ 
\pmatrix{z_1 \cr \bar{z}_2} \rightarrow U_R \, 
\pmatrix{z_1 \cr \bar{z}_2} 
\end{equation} 
Here $U_L \in \su_L$ and $U_R \in \su_R$.  We go between these two 
transformations by exchanging $\phi \leftrightarrow \psi$. 
 
We can compute the action of $\su_L$ explicitly.   Write the group 
elements as $U_L = e^{-\theta^i \, T_i}$ in terms of generators 
\begin{equation} 
T_1 = -{i \over 2} \pmatrix{0 & 1 \cr 1 & 0} ~~~;~~~ 
T_2 = {1\over 2} \pmatrix{0 & -1 \cr 1 & 0} ~~~;~~~ 
T_3 = -{i \over 2} \pmatrix{1 & 0 \cr 0 & -1} \ . 
\end{equation} 
With a little labour one can show that the infinitesimal 
transformations are explicitly realized on $(z_1,z_2)$ by the 
differential operators 
\begin{eqnarray} 
\bL_1 &=& c_\psi \, \partial_\theta + {s_\psi \over s_\theta} \, 
\partial_\phi - s_\psi \, \cot\theta \, \partial_\psi \ ,\\ 
\bL_2 &=& -s_\psi \, \partial_\theta + {c_\psi \over s_\theta} \, 
\partial_\phi  - c_\psi \, \cot\theta \, \partial_\psi \ ,\\ 
\bL_3 &=& \partial_\psi \ . 
\end{eqnarray} 
Since the exchange $(\phi \leftrightarrow \psi)$ exchanges $\su_L$ and 
$\su_R$, the $\su_R$ transformations are explicitly realized by the 
differential operators 
\begin{eqnarray} 
\bR_1 &=& c_\phi \, \partial_\theta + {s_\phi \over s_\theta} \, 
\partial_\psi - s_\phi \, \cot\theta \, \partial_\phi \ ,\\ 
\bR_2 &=& -s_\phi \, \partial_\theta + {c_\phi \over s_\theta} \, 
\partial_\psi  - c_\phi \, \cot\theta \, \partial_\phi \ , \\ 
\bR_3 &=& \partial_\phi \ . 
\end{eqnarray} 
It is  also easy to check explicitly that these operators obey the Lie 
algebra of $\su \times \su$: 
\begin{equation} 
[\bL_i,\bL_j] = \epsilon_{ijk} \, \bL_k ~~~~;~~~~ 
[\bR_{i'},\bR_{j'}] = \epsilon_{i'j'k'} \, \bR_{k'} ~~~~;~~~~ 
[\bL_i,\bR_{j'}] = 0 \ . 
\end{equation} 
The indices $i$ and $i'$ on $\bL_i$ and $\bR_{i'}$ can  be raised and
lowered freely.

\subsection{Killing vectors and vielbeins} 
$\sphere$  has six Killing vectors, which can be taken to be the 
generators of the $\su_L$ and $\su_R$ symmetries above.   That is, 
\begin{eqnarray} 
\bK_I^m &=& \bL_I^m ~~~~~~~~I = 1,2,3 \ , \\ 
      &=& \bR_{I-3}^m ~~~~~~I = 4,5,6 \ . 
\end{eqnarray} 
The corresponding one-forms have components    
\begin{eqnarray} 
\bL_{1m} &=& {\ell^2 \over 4} (c_\psi, s_\psi \, s_\theta, 0) \ , \\ 
\bL_{2m} &=& {\ell^2 \over 4} (-s_\psi, c_\psi \, s_\theta, 0) \ , \\ 
\bL_{3m} &=& {\ell^2 \over 4} (0,c_\theta,1) \ , \\ 
\bR_{1m} &=& {\ell^2 \over 4} (c_\phi, 0 , s_\phi \, s_\theta) \ , \\ 
\bR_{2m} &=& {\ell^2 \over 4} (-s_\phi, 0, c_\phi \, s_\theta)  \ , \\ 
\bR_{3m} &=& {\ell^2 \over 4} (0, 1, c_\theta) \ . 
\end{eqnarray} 
 
There are also two choices of vielbein for $\sphere$ constructed from 
the $\su_L$ and $\su_R$ generators. A vielbein is defined by  
\begin{eqnarray} 
e^a_m \, e^b_n \, \delta_{ab} &=& g_{mn}  \ ,\\ 
e^a_m \, e^b_n \, g^{mn} &=& \delta^{ab}\ . 
\end{eqnarray} 
The norm of the one-forms above is  
\begin{equation} 
\bL_{im} \, \bL_{jn} \, g^{mn} = \delta_{ij} \, {\ell^2 \over 4} 
~~~~;~~~~ 
\bR_{i'm} \, \bR_{j'n} \, g^{mn} = \delta_{i'j'} \, {\ell^2 \over 4}
\, .
\end{equation} 
Since the sphere is 3-dimensional, the $\bL$ and $\bR$ cannot of 
course be mutually orthogonal as vectors. It is readily checked that 
\begin{equation} 
\bL_{im} \, \bL_{jn} \, \delta^{ij} = g_{mn} \, {\ell^2 \over 4} \ , 
\end{equation} 
and similarly for $\bR$. Thus, we can construct a vielbein by 
identifying the group index $i$ with a tangent index $a$ and introducing
an appropriate normalization factor. The left and right vielbeins 
defined in this manner are: 
\begin{equation} 
e_{Lam} = {2 \over \ell} \bL_{am}  
~~~~~;~~~~~ 
e_{Ra'm} = - {2 \over \ell} \bR_{a'm} \ .\label{viel}  
\end{equation} 
 
\subsection{Volumes} 
In these Euler angle coordinates, the volume of the sphere is 
\begin{equation} 
{\rm Vol} = \int_0^\pi d\theta \, \int_0^{2\pi} d\phi \, \int_0^{4\pi} d\psi 
\, \sqrt{\det g} = 
\int_0^\pi d\theta \, \int_0^{2\pi} d\phi \, \int_0^{4\pi} d\psi 
\left({\ell \over 2} \right)^3 \, \sin\theta = \ell^3 \, 2\pi^2 \ .  
\end{equation} 
Accordingly, the volume form for $\sphere$ is 
\begin{equation} 
\left( {\ell \over 2}  \right)^3 \, \sin\theta \, d\theta \wedge d\phi 
\wedge d\psi 
\equiv V \epsilon_{mnr} dx^m dx^n dx^r . 
\end{equation} 
 
\subsection{Computing $N_{Ir}$ and the $\su$ projectors} 
 
The discussion of the consistent ansatz for the three-form involved a 
two-form  
\begin{equation} 
\omega = V \, \epsilon_{mnr} \, K_I^m \, dx^n \, dx^r \ , 
\end{equation} 
which is closed, and hence, on the sphere, an exact form.  So we can 
write 
\begin{equation} 
\omega = d(N_{Ir} dx^r) = \partial_n \, N_{Ir} \, dx^n \wedge dx^r 
\end{equation} 
for some $N_{Ir}$. That is, $N_{Ir}$ are defined as the solutions of  
\begin{equation} 
\partial_n N_{Ir} - \partial_r N_{In} = 2 V \epsilon_{mnr} K_I^m \ . 
\label{cond} 
\end{equation} 
It is easy to show that a solution is \footnote{We can of course add 
any closed one-form to $N_{Ir}$ and we will still have a solution; we 
will always choose to use the above solution.} 
\begin{eqnarray} 
I = 1,2,3 ~~~~~ &\Longrightarrow& N_{Im} = -\ell \, \bK_{Im}  \ ,\\ 
I = 4,5,6 ~~~~~ &\Longrightarrow& N_{Im} =  \ell \, \bK_{Im}  \ .  
\end{eqnarray} 
The defining equation (\ref{cond}) then implies 
 \begin{equation} 
\partial_n K_{Ir} - \partial_r K_{In} = {2 V \over \ell^2} 
 \epsilon_{mnr} N_I^m \ .  
\end{equation} 
We can rewrite this with tangent indices by contracting with the 
vielbein $e^m_a$, yielding  
 \begin{equation} 
\partial_a K_{Ib} - \partial_b K_{Ia} = {2  \over \ell^2} 
 \epsilon_{abc} N_I^c \ .  
\end{equation} 
Taken together with the fact that $K^I$ are Killing vectors, this 
implies 
\begin{equation} 
\nabla_a K_{Ib} = {1 \over \ell^2} \epsilon_{abc} N_I^c\ . 
\end{equation} 
 
We can construct the combinations: 
\begin{equation} 
R_{Ir} = - {N_{Ir} \over \ell} - \bK_{Im} 
~~~~~;~~~~~ 
L_{Ir} = - {N_{Ir} \over \ell} + \bK_{Im}\ . 
\end{equation} 
Clearly, 
\begin{eqnarray} 
R_{Ir} &=& 0 ~~~~~~~~~~~~~~ I=1,2,3 \ ,\\ 
       &=& - 2 \bK_{Im} = - 2 \bR_{(I-3) m} ~~~~ I=4,5,6 \ , 
\end{eqnarray} 
and 
\begin{eqnarray} 
L_{Ir} &=& 2 \bK_{Im} = 2 \bL_{Im}~~~~ I=1,2,3 \ ,\\ 
       &=& 0  ~~~~~~~~~~~~~~ I=1,2,3  \ . 
\end{eqnarray} 
Thus, these combinations act as projectors onto $\su_L$ and $\su_R$ 
respectively. In the Killing spinor equations, these projectors appear 
with flat {\it tangent} indices, i.e., $L_{Ia} = L_{Im} e^m_a$ and 
$R_{Ia} = R_{Im} e^m_a$ where $e^m_a$ is a left or right 
vielbein. Recalling the expressions for the vielbeins given in 
(\ref{viel}), 
\begin{eqnarray} 
R_{Ir} &=& 0 ~~~~~~~~~~~~~~ I=1,2,3 \ , \\ 
       &=& \ell e_{R(I-3) m} ~~~~ I=4,5,6 \ , 
\end{eqnarray} 
and 
\begin{eqnarray} 
L_{Ir} &=& \ell e_{LIm}~~~~ I=1,2,3 \ ,\\ 
       &=& 0  ~~~~~~~~~~~~~~ I=1,2,3 \ . 
\end{eqnarray} 
Since the $\su_L$ and $\su_R$ equations decouple, we can go to a 
tangent frame using $e_L$ and $e_R$ separately in each case.  So, 
choosing the left and right tangent frames in each case (call the 
indices $a$ and $a'$), we find: 
\begin{eqnarray} 
R_{Ia'} &=& 0 ~~~~~~~~~~~~~~ I=1,2,3 \ ,\\ 
       &=& \ell \delta_{(I-3)a'} ~~~~ I=4,5,6 \ , 
\end{eqnarray} 
and 
\begin{eqnarray} 
L_{Ia} &=& \ell \delta_{Ia}~~~~ I=1,2,3 \ ,\\ 
       &=& 0  ~~~~~~~~~~~~~~ I=1,2,3 \ . 
\end{eqnarray}


\end{document}